\documentclass{LMCS}

\def\dOi{9(4:15)2013}
\lmcsheading%
{\dOi}
{1--21}
{}
{}
{Feb.~10, 2013}
{Nov.~20, 2013}
{}


\ACMCCS{[{\bf Theory of computation}]: Models of
  computation---Computability; [{\bf Mathematics of computing}]:
  Mathematical analysis---Numerical analysis}

\usepackage{hyperref}
\usepackage{hyphenat}
\usepackage{amssymb}

\theoremstyle{plain}\newtheorem{example}[thm]{Example}
\theoremstyle{plain}\newtheorem{remark}[thm]{Remark}

\begin{document}

\title[APPROXIMATION SYSTEMS]{APPROXIMATION SYSTEMS FOR FUNCTIONS \\IN
  TOPOLOGICAL AND IN METRIC SPACES\rsuper*}

\author[D.~Skordev]{Dimiter Skordev}
\address{Sofia University, Faculty of Mathematics and Informatics, Sofia, Bulgaria}
\email{skordev@fmi.uni-sofia.bg}
\thanks{Research partially supported by the project ``Mathematical Logic, Algebra, and Algebraic Geometry'' of the Research Foundation -- Flanders and the Bulgarian Academy of Sciences.}

\titlecomment{{\lsuper*}A previous version of the results from Subsections \ref{S:mas} and \ref{comp} appeared in \cite{ed}}

\begin{abstract}
  \noindent A notable feature of the TTE approach to computability is the representation of the argument values and the corresponding function values by means of infinitistic names. Although convenient for the development of the theory, using such objects as data deviates from the computational practice and has a drawback from a logico-philosophical point of view. Two ways to eliminate the using of such names in certain cases are indicated in the paper. The first one is intended for the case of topological spaces with selected indexed denumerable bases. Suppose a partial function is given from one such space into another one whose selected base has a recursively enumerable index set, and suppose that the intersection of base open sets in the first space is computable in the sense of Weihrauch-Grubba. Then the ordinary TTE computability of the function is characterized by the existence of an appropriate recursively enumerable relation between indices of base sets containing the argument value and indices of base sets containing the corresponding function value. This result can be regarded as an improvement of a result of Korovina and Kudinov. The second way is applicable to metric spaces with selected indexed denumerable dense subsets. If a partial function is given from one such space into another one, then, under a semi{\hyp}computability assumption concerning these spaces, the ordinary TTE computability of the function is characterized by the existence of an appropriate recursively enumerable set of quadruples. Any of them consists of an index of element from the selected dense subset in the first space, a natural number encoding a rational bound for the distance between this element and the argument value, an index of element from the selected dense subset in the second space and a natural number encoding a rational bound for the distance between this element and the function value (in general, this set of quadruples is different from the one which straightforwardly arises from the relation used in the first way of elimination). One of the examples in the paper indicates that the computability of real functions can be characterized in a simple way by using the first way of elimination of the infinitistic names.
\end{abstract}

\maketitle

\section{Introduction}\label{S:intro}
The TTE approach to computability unifies the study of many important cases concerning functions with non-denumerable domains. A notable feature of this approach is the representation of the argument values and the corresponding function values by means of infinitistic names -- e.g. infinite sequences of natural numbers or of other constructive objects. Although convenient for the development of the theory, using such infinitistic objects as data deviates from the computational practice, where the data one processes are commonly constructive objects themselves. There is also a drawback of the approach from a logico-philosophical point of view, namely the necessity to use quantifiers over the infinitary names, which are usually an additional sort of objects different from the main sorts we are interested in. The present paper concerns some cases when the TTE computability notion can be characterized without using infinitistic names of the argument and function values, but using instead only the sort of constructive objects which can be components of the infinitistic names.

In the topological case, one usually considers two $T_0$-spaces with selected indexed bases in them. Given a partial function from the first space to the second one, the TTE computability of this function can be considered in the case when the two bases are indexed by natural numbers or possibly other constructive objects. For that purpose, the representation (called standard representation) of each of the spaces is used, at which the names of any point of the space are the enumerations of the set of all indices of base sets containing the point (cf., for instance, \cite[Definition 3.2.2]{ca}, \cite[Proposition 3.3]{cts}).\footnote{If some of the two spaces is not a $T_0$ one then the TTE definition cannot be used in its original form, but a multi-valued version of this definition is still usable (cf.\@ e.g.\@ \cite[Exercise 3.1.7]{ca} or \cite[Section~6]{mr} for such one). Although the admission of non-$T_0$ spaces does not add a generality to our considerations, we do not impose the $T_0$ requirement, because its absence causes no problems for them and imposing this requirement would create a complication in 
Remarks \ref{red_to_id} and \ref{red_to_id'}.} A possibility to avoid using infinitistic names in this situation is rather obvious, at least for the ordinary TTE computability notion concerning the transformation of the names only of the points which are in the domain of the function. Since the computability in question means the existence of an effective procedure transforming all names of the argument values into names of the corresponding function values, it is equivalent to uniform enumeration reducibility of the set of indices which corresponds to the function value to the one which corresponds to the argument value, i.e. to the existence of an enumeration operator which accomplishes the considered reduction for all points from the function domain.\footnote{The equivalence proof can be done by carrying out certain reasonings from \cite[\S 9.7]{trfec} in a more formal way (revealing also the uniformity). Let us note that computability of functions in topological spaces is actually introduced by a definition using enumeration operators in the paper \cite{ceets} (the definition is for a stronger computability notion which considers also the points in the complement of the function domain). Unfortunately, no comment is given there about the relation of the computability introduced by this definition to the one introduced by the TTE definition.}

Clearly, the above-mentioned characterization of the ordinary TTE computability by means of uniform enumeration reducibility can be stated as the requirement about the existence of a certain recursively enumerable relation between finite subsets of the set corresponding to the argument value and elements of the set corresponding to the function value.\footnote{Theorem 3.2.14 in \cite{ca} (attributed to P.\@~Hertling) can be interpreted in this way.} Under some natural additional assumptions, this requirement can be replaced by a simpler one, where the relation is already between elements of the first set and elements of the second one. Theorem 3.3 of \cite{ceets} can be regarded as a result of this kind.\footnote{Unfortunately, the formulation of that theorem needs a correction. Indeed, it may happen that $F^{-1}(\beta j)=\emptyset$ for some $j$ and some total computable function $F:X\to Y$, where $(X,\tau,\alpha)$ and $(Y,\lambda,\beta)$ satisfy the assumptions of the theorem. Then $F$ cannot satisfy condition (ii) in the conclusion of the theorem if all sets $\alpha i$ are non-empty. Here is an example of such a situation: $X=Y=\{0,1\}$, $\tau=\lambda=\{\emptyset,\{0\},\{0,1\}\}$, $\alpha 0=\beta 0=\{0\}$, $\alpha k=\beta k=\{0,1\}$ for any non-zero $k$, $F(0)=F(1)=1$, $j=0$. A possible way to correct the formulation is to additionally require that $\alpha i=\emptyset$ for some $i$ (this requirement would follow from the other assumptions in the case when $X$ has at least two different points and the topology $\tau$ is Hausdorff).\label{corr}}

The simpler binary relations mentioned above will be called approximation systems in the present paper. We give a definition of this notion encompassing also the case of bases indexed by objects different from the natural numbers. An approximation system for the function $f$ acting from a topological space $\mathbf{X}$ with selected indexed base $\{U_i\}_{i\in I}$ to a topological space $\mathbf{Y}$ with selected indexed base $\{V_j\}_{j\in J}$ 
can be described as a subset $R$ of $I\times J$ such that all statements
\begin{equation}\label{stat}
\forall x\in\mathrm{dom}(f)\big(x\in U_i\Rightarrow f(x)\in V_j\big),
\end{equation}
where $(i,j)\in R$, are correct, and the set of these statements is complete in the following sense: whenever $x\in\mathrm{dom}(f)$, $j\in J$ and $f(x)\in V_j$, some $i$ exists satisfying the conditions $(i,j)\in R$ and $x\in U_i$. In Section \ref{S:asts}, we prove that the existence of an approximation system for the function $f$ is always equivalent to its continuity, and, in the case when the indices from $I$ and~$J$ are natural numbers, the existence of a recursively enumerable approximation system for~$f$ implies its computability (defined via enumeration operators). The converse implication is proved under the assumption that the index set $J$ is recursively enumerable and,  roughly speaking, the intersection of sets from $\{U_i\}_{i\in I}$ is computable in the sense from the definition of computable $T_0$-space given in~\cite{cm} and the definition of computable topological space in \cite{ect}. Theorem~3.3 of \cite{ceets}, corrected as indicated in footnote \ref{corr}, can be obtained as a corollary from this result. An example given in Subsection \ref{S:count-ex} shows that the converse implication in question is not generally true.

The rest of the paper concerns the case, when the topologies in $\mathbf{X}$ and $\mathbf{Y}$ are generated by metrics and certain selected indexed dense subsets of the metric spaces are used in the formation of the bases. If $d$ and $\{\alpha(k)\}_{k\in K}$ are the metric and the selected indexed dense subset in $\mathbf{X}$, and $e$ and $\{\beta(l)\}_{l\in L}$ are the ones in $\mathbf{Y}$, then the indices from $I$ and $J$ encode elements of $K\times\mathbb{N}$ and $L\times\mathbb{N}$, respectively, and the statements~(\ref{stat}) have the form
$$\forall x\in\mathrm{dom}(f)\big(d(\alpha(k),x)<r_m\Rightarrow e(\beta(l),f(x))<r_n\big),$$
where $r_0,r_1,r_2,\ldots$ is a chosen computable sequence of positive rational numbers having 0 as its accumulation point. The counter{\hyp}example from Subsection \ref{S:count-ex} is adapted in Subsection~\ref{S:iUVcntas} to show that the above{\hyp}mentioned converse implication is not generally true also in this case. For the sufficient condition from Section~\ref{S:asts} ensuring the validity of this implication, a more usable assumption assuring it is given. The assumption in question is that $L$ is a recursively enumerable subset of $\mathbb{N}$ and, roughly speaking, $(X,d,\mathrm{rng}(\alpha),\alpha)$, where $X$ is the carrier of $\mathbf{X}$, is a semi{\hyp}computable metric space in the sense introduced on page 239 of \cite{ca}. More precisely, it would be required that $(X,d,\alpha)$ is a semi{\hyp}computable metric space in the sense of the following definition: a {\em semi{\hyp}computable metric space} is any triple $(Z,r,\gamma)$ such that $(Z,r)$ is a metric space, $\gamma$ is a partial mapping of $\mathbb{N}$ into $Z$, the set $\mathrm{rng}(\gamma)$ is dense in the metric space $(Z,r)$ and the set
$$\left\{\,(l,l^\prime,p,q)\ \left|\ l,l^\prime\in\mathrm{dom}(\gamma),\ p,q\in\mathbb{N}\setminus\{0\},\ r(\gamma(l),\gamma(l^\prime))<\frac{p}{q}\right.\right\}$$
is recursively enumerable.\footnote{This definition of the notion will be taken for granted throughout the paper. In \cite[Definition 6.1]{crts}, the term ``upper semi{\hyp}computable metric space'' is used for a similar, but slightly different notion. Up to inessential details, the difference is that one requires there $\mathrm{dom}(\gamma)$ to be recursive, whereas only the recursive enumerability of $\mathrm{dom}(\gamma)$ is implied by the definition we accept here. The effective metric spaces in the sense of \cite{emsrr} are surely semi{\hyp}computable metric spaces in the sense accepted here.} Subsection~\ref{S:iUVcntas} ends with an example which gives a simple characterization of the computability of real functions.

Another kind of approximation systems, which makes sense only in the case of metric spaces, is introduced and studied in Subsections \ref{S:mas} and \ref{comp}, namely a different kind of completeness of the considered set of statements is required. This notion can be regarded as a slight generalization of a notion introduced in \cite{ed}. We think its idea is considerably closer to the computational practice than the idea of the above{\hyp}mentioned notion.\footnote{Examples showing the relation of this other kind of approximation systems to approximate computation of real functions can be found in \cite{ed}.} Results similar to the ones from the previous sections are proved for it, but the TTE computability referred to corresponds now to the representation of the spaces as metric ones (i.e. the points are named by certain sequences of indices of approximating elements from the corresponding dense subsets).

It is a known fact that the topological computability in metric spaces, which is considered in Subsection \ref{S:iUVcntas}, and the metric one considered in Subsection \ref{comp} are equivalent under certain not very restrictive assumptions (cf., for instance, \cite[Theorem 8.1.4]{ca}). For the sake of completeness, this subject is investigated in some detail in Subsection \ref{alphabeta-UV}. It begins with two examples showing that none of these two computabilities implies the other one in the most general case, and then the two notions are shown to be equivalent in the case of semi{\hyp}computable metric spaces.\footnote{The lemma used for proving this equivalence statement is present, in essence, also in the paper \cite{crts}, but the last one appeared when the revised version of the present paper was already submitted to the referees.}

Subsection \ref{meto} is devoted to two constructions which transform one kind of approximation systems for functions in metric spaces into the other kind. The first of these constructions transforms approximation systems of the kind from Subsection \ref{S:tasms} into ones of the kind from Subsection \ref{S:mas}, and the second one works in the opposite direction. It is indicated that recursive enumerability is preserved by these constructions under certain natural assumptions.

\section{Approximation systems for functions in topological spaces}\label{S:asts}
In this section, we suppose that $\mathbf{X}$ and $\mathbf{Y}$ are topological spaces with carriers $X$ and $Y$, respectively, ${\mathcal U}=\left\{U_i\right\}_{i\in I}$ is an indexed base of $\mathbf{X}$, ${\mathcal V}=\left\{V_j\right\}_{j\in J}$ is an indexed base of $\mathbf{Y}$, and $f$ is a function from $E$ to $Y$, where $E\subseteq X$.

 \subsection{Definition of the notion; its connection to continuity} 

\begin{defi}\label{D:UVas}
A {\em$({\mathcal U},{\mathcal V})${\hyp}approximation system} for $f$ is a subset $R$ of the Cartesian product $I\times J$ such that
\begin{equation}\label{E:UVas}
\forall x\in E\ \forall j\in J\,\big(f(x)\in V_j\Leftrightarrow\exists i\,\big((i,j)\in R\ \&\ x\in U_i\big).
\end{equation}
\end{defi}

\begin{remark}{\em The condition (\ref{E:UVas}) is equivalent to
\begin{equation}\label{E:f^-1}
\forall j\in J\left(f^{-1}(V_j)=\bigcup\left\{\,U_i\cap E\,\left|\,(i,j)\in R\right.\right\}\right).
\end{equation}
If $E=X$ then the above condition reduces to the simpler one
\begin{equation}\label{E:f^-1t}
\forall j\in J\left(f^{-1}(V_j)=\bigcup\left\{\,U_i\,\left|\,(i,j)\in R\right.\right\}\right).
\end{equation}
The condition (\ref{E:UVas}) is equivalent also to one formulated in terms of composition of relations. Let ${\mathcal U}^{-1}=\{(x,i)|i\in I\,\&\,x\in U_i\}$, ${\mathcal V}^{-1}=\{(y,j)|j\in J\,\&\,y\in V_j\}$. If we identify the functions $f$ and $\mathrm{id}_E$ with the sets $\{(x,f(x))|x\in E\}$ and $\{(x,x)|x\in E\}$, respectively, then (\ref{E:UVas}) is equivalent to the equality $f\circ{\mathcal V}^{-1}=\mathrm{id}_E\circ{\mathcal U}^{-1}\circ R$, where $\circ$ denotes left-to-right composition. This equality reduces to $f\circ{\mathcal V}^{-1}={\mathcal U}^{-1}\circ R$ in the case of $E=X$. 
}\end{remark}

\begin{prop}\label{P:ex_as<->cntns}
A $({\mathcal U},{\mathcal V})${\hyp}approximation system for $f$ exists iff $f$ is continuous.
\end{prop}

\proof\hfill

$(\Rightarrow)$ Let $R$ be a ${\mathcal U},\!{\mathcal V}${\hyp}approximation system for $f$. To prove that $f$ is continuous, suppose $j\in J$, $x\in f^{-1}(V_j)$. Then, by \eqref{E:f^-1}, there exists some $i$ in $I$ such that $x\in U_i\cap E\subseteq f^{-1}(V_j)$.

$(\Leftarrow)$ Let $f$ be continuous. We set $R=\left\{(i,j)\in I\times J\,\left|\,U_i\cap E\subseteq f^{-1}(V_j)\right.\right\}$. To show that $R$ is a ${\mathcal U},\!{\mathcal V}${\hyp}approximation system for $f$, suppose $x\in E$ and $j\in J$. If $f(x)\in V_j$ then $x\in f^{-1}(V_j)$, hence (thanks to the continuity of $f$) $x\in U_i$ and $U_i\cap E\subseteq f^{-1}(V_j)$ for some $i\in I$, and clearly $(i,j)\in R$ for this $i$. Conversely, if $(i,j)\in R$ and $x\in U_i$ for some $i$, then $x\in U_i\cap E$ and therefore $x\in f^{-1}(V_j)$, hence $f(x)\in V_j$.\qed

\begin{remark}\label{maxUV}
{\em If $f$ is continuous then the set $\left\{(i,j)\in I\times J\,\left|\,U_i\cap E\subseteq f^{-1}(V_j)\right.\right\}$ considered in the above proof is the maximal $({\mathcal U},{\mathcal V})${\hyp}approximation system for $f$ in the sense that it contains as subsets all other ones.}
\end{remark}

\begin{remark}\label{red_to_id}{\em
Let us set ${\mathcal U}\upharpoonright E=\{U_i\cap E\}_{i\in I}$, $f^{-1}(\mathcal V)=\left\{f^{-1}(V_j)\right\}_{j\in J}$. Then ${\mathcal U}\upharpoonright E$ and $f^{-1}(\mathcal V)$ are indexed bases for two topologies on $E$. Let $\mathbb{X^\prime}$ and $\mathbb{Y^\prime}$ be the corresponding two topological spaces with carrier $E$. The mapping $\mathrm{id}_E$ can be regarded as a function from $\mathbb{X^\prime}$ to $\mathbb{Y^\prime}$. It is immediately seen that a subset of $I\times J$ is a $({\mathcal U},{\mathcal V})${\hyp}approximation system for $f$ iff this subset is a $({\mathcal U}\upharpoonright E,f^{-1}(\mathcal V))${\hyp}approximation system for $\mathrm{id}_E$.
}\end{remark} 

 \subsection{Approximation systems and computability in topological spaces}\label{S:ascts}
In this subsection, we additionally suppose that $I\subseteq\mathbb{N}$ and $J\subseteq\mathbb{N}$.

\begin{defi}\label{D:UVcomp}
For any $x\in X$ and any $y\in Y$, we set
$$[x]_{\mathcal U}=\{i\in I|x\in U_i\},\ \ [y]_{\mathcal V}=\{j\in J|y\in V_j\},$$ respectively.\footnote{Since we do not assume that $\mathbf{X}$ and $\mathbf{Y}$ are necessarily $T_0$-spaces, the mappings $x\mapsto[x]_{\mathcal U}$ and $y\mapsto[y]_{\mathcal V}$ are not necessarily injective.} The function $f$ will be called {\em$({\mathcal U},{\mathcal V})${\hyp}computable}, if an enumeration operator~$F$ exists such that $[f(x)]_{\mathcal V}=F([x]_{\mathcal U})$ for any $x\in E$.
\end{defi}

\begin{remark}
{\em As indicated in Section \ref{S:intro}, the above definition is equivalent to a TTE-style one which uses the enumerations of $[x]_{\mathcal U}$ and of $[f(x)]_{\mathcal V}$ as names of $x$ and of $f(x)$, respectively.}
\end{remark}

\begin{remark}\label{red_to_id'}{\em
Let ${\mathcal U}\upharpoonright E$, $f^{-1}(\mathcal V)$, $\mathbb{X^\prime}$ and $\mathbb{Y^\prime}$ be defined as in Remark \ref{red_to_id}. Since $[x]_{\mathcal U}=[x]_{{\mathcal U}\upharpoonright E}$ and $[f(x)]_{\mathcal V}=[x]_{f^{-1}(\mathcal V)}$ for any $x$ in~$E$, the $({\mathcal U},{\mathcal V})${\hyp}computability of $f$ turns equivalent to the $({\mathcal U}\upharpoonright E,f^{-1}(\mathcal V))${\hyp}computability of~$\mathrm{id}_E$.
}\end{remark}

\begin{thm}\label{T:exreas->UVcomp}
If there exists a recursively enumerable $\,({\mathcal U},{\mathcal V})${\hyp}approximation system for~$f$ then $f$ is $\,({\mathcal U},{\mathcal V})${\hyp}computable.
\end{thm}

\proof Let $R$ be a recursively enumerable $({\mathcal U},{\mathcal V})${\hyp}approximation system for $f$. The mapping $F:{\mathcal P}(\mathbb{N})\to{\mathcal P}(\mathbb{N})$ defined by
$$F(M)=\left\{\,j\,\left|\,\exists i\in M\,\big((i,j)\in R\big)\right.\right\}$$
is an enumeration operator. If $x$ is an arbitrary element of $E$ then $[f(x)]_{\mathcal V}=F([x]_{\mathcal U})$, since for any $j$ the following equivalences hold:
\begin{multline*}
j\in[f(x)]_{\mathcal V}\Leftrightarrow j\in J\ \&\ f(x)\in V_j\Leftrightarrow\exists i\,\big((i,j)\in R\ \&\ x\in U_i)\Leftrightarrow\\
\Leftrightarrow\exists i\in[x]_{\mathcal U}\,\big((i,j)\in R)\Leftrightarrow j\in F([x]_{\mathcal U}).\ \ \ 
\end{multline*}

\vspace{-6.5mm}\qed

\medskip

\noindent The converse theorem is not generally true (a counter-example to it will be given in the next section). However, it is true if the set $J$ is recursively enumerable and the indexed base ${\mathcal U}$ satisfies a certain additional requirement, which essentially coincides with the main requirement in the definition of computable $T_0$-space given in \cite{cm} and the definition of computable topological space in \cite{ect} (cf. \cite[(2) on page 349]{cm} and \cite[(10) on page 1387]{ect}).\footnote{This requirement is surely satisfied in the case when the topological space $\mathbf{X}$ considered together with the indexed base $\mathcal U$ is effectively enumerable in the sense of \cite{ceets}.}

\begin{thm}\label{T:UVcomp->exreas}
Let the set $J$ be recursively enumerable, and let a recursively enumerable subset $H$ of $I^3$ exists such that
\begin{equation}\label{E:compmeet}
\forall i_1,i_2\in I\,\left(U_{i_1}\cap U_{i_2}=\bigcup\big\{U_i\,|\,(i_1,i_2,i)\in H\big\}\right).
\end{equation}
If $f$ is $\,({\mathcal U},{\mathcal V})${\hyp}computable then there exists a recursively enumerable $\,({\mathcal U},{\mathcal V})${\hyp}approximation system for $f$.
\end{thm}

\proof
For $k\!=\!2,3,4,\ldots$, we define a set $H_k$ by means of the equalities
$$H_2=H,\ \ H_{k+1}=\left\{(i_1,\ldots,i_k,i_{k+1},i)\left|\,\exists i^\prime\big((i_1,\ldots,i_k,i^\prime)\in H_k\ \&\ (i^\prime,i_{k+1},i)\in H\big)\right.\right\}.$$
One proves by induction that, for any integer greater $k$ than 1, the set $H_k$ is a subset of~$I^{k+1}$ such that
$$\forall i_1,\ldots,i_k\in I\,\big(U_{i_1}\cap\ldots\cap U_{i_k}=\bigcup\{U_i\,|\,(i_1,\ldots,i_k,i)\in H_k\}\big).$$
Besides, $\bigcup_{k=2}^\infty H_k$ is a recursively enumerable subset of the set of all finite sequences of natural numbers. We additionally set $I_0=\left\{i\,\left|\,\exists i^\prime\big((i^\prime,i^\prime,i)\in H\big)\right.\right\}$. Then $I_0\subseteq I$, $I_0$ is recursively enumerable, and, by applying (\ref{E:compmeet}) with $i_1=i_2=i^\prime$, we see that $$\forall i^\prime\in I\,\forall x\in U_{i^\prime}\,\exists i\in I_0\big(x\in U_i\subseteq U_{i^\prime}\big),$$ hence each element of $X$ belongs to some $U_i$ with $i\in I_0$.\footnote{Another way to construct a recursively enumerable subset $I_0$ of $I$ with this property is by setting $I_0=\left\{i\,\left|\,\exists i^\prime\big((i,i, i^\prime)\in H\big)\right.\right\}$.} Let $F$ be an enumeration operator with the property from Definition \ref{D:UVcomp}, and $R$ be the set of all $(i,j)$ such that $i\in I_0$ and $j\in F(\emptyset)\cap J$, or $(i_1,\ldots,i_k,i)\in H_k$ and $j\in F(\{i_1,\ldots,i_k\})\cap J$ for some natural number $k$ greater than 1 and some $i_1,\ldots,i_k$. Clearly, $R$ is recursively enumerable. We will show that $R$ is a $({\mathcal U},{\mathcal V})${\hyp}approximation system for $f$. Of course, $R\subseteq I\times J$. Now let $x\in E$, $j\in J$.

Suppose first that $f(x)\in V_j$. Then $j\in[f(x)]_{\mathcal V}$, therefore $j\in F([x]_{\mathcal U})$, hence $j\in F(M)$ for some finite subset $M$ of $[x]_{\mathcal U}$. If $M=\emptyset$, then we can satisfy the conditions $(i,j)\in R$ and $x\in U_i$ by taking an arbitrary $i\in I_0$ such that $x\in U_i$. Let now $M=\{i_1,\ldots,i_k\}$, where $k>0$ and $i_1,\ldots,i_k\in[x]_{\mathcal U}$, hence $x\in U_{i_1}\cap\ldots\cap  U_{i_k}$. Without loss of generality (since  $i_1,\ldots,i_k$ need not necessarily be pairwise different), we may assume that $k\ge 2$. Then $x\in U_i$ for some $i$ satisfying the condition $(i_1,\ldots,i_k,i)\in H_k$. We will have again $(i,j)\in R$  and $x\in U_i$ for this $i$.

Suppose now that $(i,j)\!\in\!R$ and $x\!\in\!U_i$ for some $i$. If $j\in F(\emptyset)$ then \mbox{$j\in F([x]_{\mathcal U})=[f(x)]_{\mathcal V}$,} thus $f(x)\in V_j$. And if $(i_1,\ldots,i_k,i)\in H_k$,  $j\in F(\{i_1,\ldots,i_k\})$, where $k>1$, then \mbox{$i_1,\ldots,i_k\in I$} and $U_i\!\subseteq\!U_{i_1}\cap\ldots\cap  U_{i_k}$, hence $x\in U_{i_1}\cap\ldots\cap  U_{i_k}$, and therefore \mbox{$\{i_1,\ldots,i_k\}\!\subseteq\![x]_{\mathcal U}$.} Thus again $j\in F([x]_{\mathcal U})=[f(x)]_{\mathcal V}$, and $f(x)\in V_j$.\qed

\begin{cor}\label{C:UVcomp<->exreas}
Under the assumptions of Theorem \ref{T:UVcomp->exreas}, the function $f$ is $\,({\mathcal U},{\mathcal V})${\hyp}computable iff there exists a recursively enumerable $\,({\mathcal U},{\mathcal V})${\hyp}approximation system for $f$.
\end{cor}

\begin{cor}\label{C:UVcomp<->exRf^-1t}
Let $E=X$, the set $J$ be recursively enumerable, and let a recursively enumerable subset $H$ of $I^3$ exist which satisfies the condition (\ref{E:compmeet}). Then $f$ is $\,({\mathcal U},{\mathcal V})${\hyp}computable iff a recursively enumerable subset $R$ of $\mathbb{N}^2$ exists which satisfies the condition~(\ref{E:f^-1t}).
\end{cor}

Theorem 3.3 of \cite{ceets} (corrected as indicated in footnote \ref{corr}) can be derived from the particular instance of the above corollary when $I=J=\mathbb{N}$ and $U_i=\emptyset$ for some $i\in I$. Indeed, let $(X,\tau,\alpha)$ and $(Y,\lambda,\beta)$ be effectively enumerable topological spaces in the sense of that paper, $\alpha i$ be empty for some $i$, $f:X\to Y$ be a total function. Let us set $I=J=\mathbb{N}$, $U_i=\alpha i$, $V_j=\beta j$. Then the computability of $f$ in the sense of \cite{ceets} will be equivalent to its $({\mathcal U,\mathcal V}) ${\hyp}computability. By \cite[condition (ii) in Definition 2.1]{ceets}, a recursive function~$g$ exists such that
$$U_{i_1}\cap U_{i_2}=\bigcup_{n\in\mathbb{N}}U_{g(i_1,i_2,n)}$$
for all $i_1,i_2\in\mathbb{N}$. If we set
$$H=\{(i_1,i_1,g(i_1,i_2,n))\,|\,i_1,i_2,n\in\mathbb{N}\},$$
then (\ref{E:compmeet}) holds. Therefore, by Corollary \ref{C:UVcomp<->exRf^-1t}, the computability of $f$ in the sense of \cite{ceets} will be equivalent to the existence of a recursively enumerable subset $R$ of $\mathbb{N}^2$ which satisfies the condition (\ref{E:f^-1t}), i.e.
\begin{equation}\label{alphaR}
\forall j\in\mathbb{N}\left(f^{-1}(\beta j)=\bigcup\{\alpha i\,|\,(i,j)\in R\}\right).
\end{equation}
We will now show that the existence of such a set $R$ is equivalent to condition (ii) of Theorem~3.3 in \cite{ceets} for the function $f$, i.e. to the existence of two-argument recursive function $h$ satisfying
\begin{equation}\label{E:h}
\forall j\in\mathbb{N}\left(f^{-1}(\beta j)=\bigcup_{t\in\mathbb{N}}\alpha h(t,j)\right).
\end{equation}
Suppose $R$ is a recursively enumerable subset of $\mathbb{N}^2$ satisfying (\ref{alphaR}). Let $i_0$ be a natural number such that $\alpha i_0=\emptyset$, $\chi$ be a 3{\hyp}argument primitive recursive function such that $$\forall i,j\in\mathbb{N}\,\big((i,j)\in R\Leftrightarrow\exists s\in\mathbb{N}\,\big(\chi(i,j,s)=0\big)\big),$$ and $\pi_1,\pi_2$ be unary primitive recursive functions such that $\{(\pi_1(t),\pi_2(t))\,|\,t\in\mathbb{N}\}=\mathbb{N}^2$. If we set
$$h(t,j)=\left\{\begin{array}{ll}\pi_1(t)&\textnormal{if $\chi(\pi_1(t),j,\pi_2(t))=0$,}\\
i_0&\textnormal{otherwise,}\end{array}\right.$$
then $h$ is a primitive recursive function satisfying (\ref{E:h}). Conversely, if $h$ is a two-argument recursive function satisfying (\ref{E:h}), and we set $R=\{(h(t,j),j)\,|\,t\in\mathbb{N}\}$, then $R$ is a recursively enumerable set satisfying (\ref{alphaR}).

 \subsection{A counter-example to the converse theorem of Theorem 2.9}\label{S:count-ex}
To construct the counter-example in question, we will partition $\mathbb{N}$ into disjoint two-elements sets $P_0$, $P_1$, $P_2$, $\ldots\,$ such that:
\begin{enumerate}[label=\({\alph*}]
\item for any $x\in\mathbb{N}$, the two numbers in $P_x$ differ by $x+1$;
\item no recursive function exists whose value belongs to $P_x$ for any $x\in\mathbb{N}$.
\end{enumerate}
Let $k_0,k_1,k_2,\ldots$ be a sequence of natural numberså which is dominated by no recursive function, and, in addition, let $k_{l+1}>k_l+2l+1$ for any $l\in\mathbb{N}$. We construct subsets $C_0,C_1,C_2,\ldots$ of $\mathbb{N}^2$ in such a way that $C_0=\{(k_l,k_l+2l+1)\,|\,l\in\mathbb{N}\}$ and, for any $r\in\mathbb{N}$, we have 
$C_{r+1}=C_r\cup\{(m_r,n_r)\}$, where:
\begin{enumerate}[label=(\roman*)]
\item $m_r$ and $n_r$ are natural numbers occurring in no pair from $C_r$;
\item $m_r<n_r$, and $n_r-m_r\ne n-m$, whenever $(m,n)\in C_r$;
\item if $r$ is even then $m_r$ is the least of the natural numbers occurring in no pair from~$C_r$;
\item if $r$ is odd then $n_r-m_r$ is the least positive integer which is different from $n-m$, whenever $(m,n)\in C_r$.
\end{enumerate}
Then, for any $x\in\mathbb{N}$, there exists exactly one pair $(m,n)$ in $\bigcup_{r=0}^\infty C_r$ with $n-m=x+1$. We take as $P_x$ the set of the members of this pair.

Let $X=Y=I=J=\mathbb{N}$. For any $x\in\mathbb{N}$, let $U_i=\{x\}$ for $i\in P_x$, $V_x=\{x\}$ (thus both $\mathbf{X}$ and $\mathbf{Y}$ are the discrete topological space with carrier $\mathbb{N}$). Then the function $\mathrm{id}_{\mathbb{N}}$ is $({\mathcal U},{\mathcal V})${\hyp}computable, since $[\mathrm{id}_{\mathbb{N}}(x)]_{\mathcal V}=\{x\}\!=\!F(P_x)=F([x]_{\mathcal U})$, where $F:{\mathcal P}(\mathbb{N})\to{\mathcal P}(\mathbb{N})$ is defined by means of the equality $F(M)=\{i^\prime-i-1\,|\,i,i^\prime\in M,i^\prime>i\}$. Suppose $R$ is a recursively enumerable $({\mathcal U},{\mathcal V})${\hyp}approximation system for $\mathrm{id}_{\mathbb{N}}$. Then
\begin{equation}\label{E:x=j<->exiinP_x}
\forall x,j\in\mathbb{N}\big(x=j\Leftrightarrow\exists i\in P_x\big((i,j)\in R\big)\big).
\end{equation}
It follows from here that $\forall j\in\mathbb{N}\,\exists i\big((i,j)\in R\big)$.
By the recursive enumerability of $R$, this implies the existence of a recursive function $\iota$ such that $(\iota(j),j)\in R$ for any $j\in\mathbb{N}$. Let $j$ be an arbitrary natural number. Since any number from $\mathbb{N}$ belongs to some of the sets $P_0,P_1,P_2,\ldots$, there exists some $x\in\mathbb{N}$ such that $\iota(j)\in P_x$. By (\ref{E:x=j<->exiinP_x}), we conclude from here that $x=j$ and consequently $\iota(j)\in P_j$. Since this will hold for any $j\in\mathbb{N}$, we get a contradiction. Hence no recursively enumerable $({\mathcal U},{\mathcal V})${\hyp}approximation system for $\mathrm{id}_{\mathbb{N}}$ exists.

The above counter-example shows that, even in the case when $I=J=\mathbb{N}$ and \mbox{$E=X$,} one cannot omit the assumption in Theorem \ref{T:UVcomp->exreas} about the existence of a recursively enumerable set $H$ with the property (\ref{E:compmeet}). Thus it is not generally true in this case that a recursively enumerable subset $R$ of $\mathbb{N}^2$ with the property (\ref{E:f^-1t}) exists, whenever $f$ is $({\mathcal U},{\mathcal V})${\hyp}computable.

\section{Approximation systems for functions in metric spaces}
In this section, in addition to the assumptions made in the beginning of Section \ref{S:asts}, we will suppose that the topologies of $\mathbf{X}$ and $\mathbf{Y}$ are generated by a metric $d$ in $X$ and a metric $e$ in $Y$, respectively. As to the indexed bases $\mathcal U$ and $\mathcal V$, they will be supposed to be defined in the following way. Some index sets $K$, $L$ and mappings $\alpha:K\to X$, $\beta:L\to Y$ are supposed to be given such that $\mathrm{rng}(\alpha)$ is dense in $X$ and $\mathrm{rng}(\beta)$ is dense in $Y$, as well as some bijections $\kappa:K\times\mathbb{N}\to I$, $\lambda:L\times\mathbb{N}\to J$ and a computable sequence $r_0,r_1,r_2,\ldots$ of positive rational numbers such that 0 is an accumulation point of it.\footnote{Certain typical choices for the sequence $r_0,r_1,r_2,\ldots$ are $1,\frac{1}{2},\frac{1}{2^2},\frac{1}{2^3},\ldots$ and $1,\frac{1}{2},\frac{1}{3},\frac{1}{4},\ldots$ or, say, the sequence $\frac{1}{1},\frac{1}{2},\frac{2}{1},\frac{1}{3},\frac{2}{2},\frac{3}{1},\frac{1}{4},\frac{2}{3},\frac{3}{2},\frac{4}{1},\ldots$ enumerating all positive rational numbers.} Then we set $U_{\kappa(k,m)}=B_d(\alpha(k),r_m)$ for all \mbox{$(k,m)\in K\times\mathbb{N}$,} \mbox{$V_{\lambda(l,n)}=B_e(\beta(l),r_n)$} for all $(l,n)\in L\times\mathbb{N}$, where, for any positive real number $r$, $B_d(x,r)=\{\bar{x}\in X|\,d(x,\bar{x})<r\}$ and $B_e(y,r)=\{\bar{y}\in Y|\,e(y,\bar{y})<r\}$ for any $x\in X$ and for any $y\in Y$, respectively.

 \subsection{Topological approximation systems for functions in metric spaces}\label{S:tasms}
Definition~\ref{D:UVas} yields the following statement: a subset $R$ of $I\times J$ is a $({\mathcal U},{\mathcal V})${\hyp}approximation system for $f$ iff
\begin{multline*}
\forall x\in E\ \forall(l,n)\in L\times\mathbb{N}\,\big(e(\beta(l),f(x))<r_n\Leftrightarrow\\
\exists(k,m)\in K\times\mathbb{N}\big((\kappa(k,m),\lambda(l,n))\in R\ \&\ d(\alpha(k),x)<r_m\big)\big).
\end{multline*}
If $R\subseteq I\times J$, and we set $S=\{(k,m,l,n)\in K\times\mathbb{N}\times L\times\mathbb{N}\,|\,(\kappa(k,m),\lambda(l,n))\in R\}$, then the above condition is equivalent to the following one:
\begin{multline}\label{E:tasms}
\forall x\in E\ \forall(l,n)\in L\times\mathbb{N}\,\big(e(\beta(l),f(x))<r_n\Leftrightarrow\\
\exists k,m\big((k,m,l,n)\in S\ \&\ d(\alpha(k),x)<r_m\big)\big).
\end{multline}
Evidently, there is a one-to-one correspondence between the $({\mathcal U},{\mathcal V})${\hyp}approximation systems for~$f$ and the sets $S\subseteq K\times\mathbb{N}\times L\times\mathbb{N}$ which satisfy (\ref{E:tasms}). Using such sets has the advantage of eliminating the mappings $\kappa$ and $\lambda$. Therefore we give the next definition.

\begin{defi}\label{D:alphabetaas}
A {\em topological $(\alpha,\beta)${\hyp}approximation system} for $f$ is any subset $S$ of the Cartesian product $K\times\mathbb{N}\times L\times\mathbb{N}$ such that (\ref{E:tasms}) holds.
\end{defi}

Proposition \ref{P:ex_as<->cntns} yields the next statement.

\begin{cor}\label{tas<->cont}
A topological $(\alpha,\beta)${\hyp}approximation system for $f$ exists iff $f$ is continuous.
\end{cor}

Of course, the notion of a topological $(\alpha,\beta)${\hyp}approximation system for $f$ depends (except for some specific cases) also on the choice of the sequence $r_0,r_1, r_2,\ldots$ However, the existence of such a system does not depend on this choice, as seen from Corollary \ref{tas<->cont}.

\begin{remark}\label{maxtop}
{\em Taking into account Remark \ref{maxUV}, it is easy to see that the set
\begin{equation}\label{max}
\left\{(k,l,m,n)\in K\times\mathbb{N}\times L\times\mathbb{N}\,\left|\,\forall x\in E\,\big(d(\alpha(k),x)<r_m\Rightarrow e(\beta(l),f(x))<r_n\big)\right.\right\}
\end{equation}
is the maximal topological $(\alpha,\beta)${\hyp}approximation system for $f$.}
\end{remark}

 \subsection{The interconnection between \texorpdfstring{$({\mathcal U},{\mathcal V})$}{(U,V)}{\hyp}computability and topological \texorpdfstring{$(\alpha,\beta)$}{(alpha,beta)}{\hyp}approximation systems}\label{S:iUVcntas}
In this subsection, we additionally suppose that $I$, $J$, $K$, $L$ are subsets of $\mathbb{N}$, and that $\kappa$ and $\lambda$ are restrictions of some computable bijection from $\mathbb{N}^2$ to $\mathbb{N}$, whose value at the pair $(s,t)$ will be denoted by $\langle s,t\rangle$.

\begin{thm}\label{T:exrealphabetaas->UVcomp}
If there exists a recursively enumerable topological $(\alpha,\beta)${\hyp}approximation system for $f$ then $f$ is $\,({\mathcal U},{\mathcal V})${\hyp}computable.
\end{thm}

\proof Let $S$ be a recursively enumerable topological $(\alpha,\beta)${\hyp}approximation system for $f$. Then Theorem~\ref{T:exreas->UVcomp} can be applied, since the set $R=\{\langle k,m\rangle,\langle l,n\rangle|(k,m,l,n)\in S\}$ is a recursively enumerable $\,({\mathcal U},{\mathcal V})${\hyp}approximation system for $f$.\qed

The converse of Theorem \ref{T:exrealphabetaas->UVcomp} is not generally true. 

\begin{example}\label{no_eff}
{\em Let $\mathbf{X},\mathbf{Y},I,J$ and the two-elements sets $P_x$ be the same as in Subsection~\ref{S:count-ex}. The considered topology can be generated by the usual metric in $\mathbb{N}$, namely the absolute value of the distance. We take each of the metrics $d$ and $e$ to be this one. Let $K=L=\mathbb{N}$, $\alpha(k)=x$ for any $x\in\mathbb{N}$ and each $k\in P_x$, $\beta(l)=l$ for any $l\in\mathbb{N}$. Let none of the numbers $r_0,r_1,r_2,\ldots$ be greater than 1. Then, whenever $x\in\mathbb{N}$ and $k\in P_x$, $U_{\langle k,m\rangle}=\{x\}$ for all $m\in\mathbb{N}$, and $V_{\langle l,n\rangle}=\{l\}$ for all $l,n\in\mathbb{N}$. Therefore
$[x]_{\mathcal U}=\{\langle k,m\rangle|k\in P_x\ \&\ m\in\mathbb{N}\}$, $[x]_{\mathcal V}=\{\langle x,n\rangle|n\in\mathbb{N}\}$ for all $x\in\mathbb{N}$. Let us define 
$F:{\mathcal P}(\mathbb{N})\to{\mathcal P}(\mathbb{N})$ by the equality
$$F(M)=\{\langle k^\prime-k-1,n\rangle|\,k,k^\prime,n\in\mathbb{N}\ \&\ \langle k,0\rangle,\langle k^\prime,0\rangle\in M\ \&\ k^\prime>k\}.$$
Then $F$ is an enumeration operator, and $[x]_{\mathcal V}=F([x]_{\mathcal U})$ for any $x\in\mathbb{N}$, hence the function $\mathrm{id}_\mathbb{N}$ 
is $\,({\mathcal U},{\mathcal V})${\hyp}computable. Suppose there exists a recursively enumerable topological $(\alpha,\beta)${\hyp}approximation system $S$ for $\mathrm{id}_\mathbb{N}$. Then, by (\ref{E:tasms}),
$$\forall x,l\in\mathbb{N}\big(l=x\Leftrightarrow\exists(k,m)\in\mathbb{N}^2\big((k,m,l,0)\in S\ \&\ \alpha(k)=x\big)\big).$$
Hence,
$$\forall x,l\in\mathbb{N}\big(x=l\Leftrightarrow\exists k\in P_x\big((k,l)\in R\big)\big),$$
where $R=\left\{(k,l)\left|\exists m\big((k,m,l,0)\in S\big)\right.\right\}$. Since $R$ is recursively enumerable, we get a contradiction from here in the same way as we got one from (\ref{E:x=j<->exiinP_x}).}
\end{example}

Theorem \ref{T:UVcomp->exreas} yields the following sufficient condition for the validity of the implication from the $\,({\mathcal U},{\mathcal V})${\hyp}computability of $f$ to the existence of a recursively enumerable topological $(\alpha,\beta)${\hyp}approximation system for $f$.

\begin{thm}\label{T:exrealphabetaas<-UVcomp}
Let the set $L$ be recursively enumerable, and let a recursively enumerable subset $H$ of $K\times\mathbb{N}\times K\times\mathbb{N}\times K\times\mathbb{N}$ exists such that
\begin{equation*}
\forall k_1,k_2\in K\,\forall m_1,m_2\in\mathbb{N} \left(U_{\langle k_1,m_1\rangle}\cap U_{\langle k_2,m_2\rangle}=\bigcup\left\{U_{\langle k,m\rangle}\,\left|\,(k_1,m_1,k_2,m_2,k,m)\in H\right.\right\}\right).
\end{equation*}
If $f$ is $\,({\mathcal U},{\mathcal V})${\hyp}computable then there exists a recursively enumerable topological $(\alpha,\beta)${\hyp}approximation system for $f$.
\end{thm}

To prove a more usable sufficient condition, we now introduce a strict partial ordering $<_d$ in $X\times\mathbb{N}$ as follows: if $(x,m),(x^\prime,m^\prime)\in X\times\mathbb{N}$ then $(x,m)<_d(x^\prime,m^\prime)$ iff $d(x^\prime,x)<r_{m^\prime}-r_m$ (the irreflexivity of this relation is evident, and its transitivity is easily verifiable). The statement (a) of the lemma below suggests an interpretation of the interrelation \mbox{$(x,m)<_d(x^\prime,m^\prime)$} as a certain intensional kind of inclusion of the closed ball \mbox{$\overline{B}_d(x,r_m)=\{\bar{x}\in X|d(x,\bar{x})\le r_m\}$} in the open ball $B_d(x^\prime,r_{m^\prime})$.\footnote{The mentioned kind of inclusion becomes an extensional one if $\mathbf{X}$ is an Euclidean space, because then \mbox{$(x,m)<_d(x^\prime,m^\prime)$} holds iff $\overline{B}_d(x,r_m)\subseteq B_d(x^\prime,r_{m^\prime})$. The interrelation $(\alpha(k),m)<_d(\alpha(k^\prime),m^\prime)$ coincides (in the case when $r_m=2^{-m}$ for any $m\in\mathbb{N}$) with the formal inclusion introduced in \cite[Section 2]{cms}.}

\begin{lem}\label{L:forminclprop}\hfill
\begin{enumerate}[label=\({\alph*}]
\item $\forall(x,m),(x^\prime,m^\prime)\!\in\!X\!\times\!\mathbb{N}\,\left((x,m)\!<_d\!(x^\prime,m^\prime)\Rightarrow\!\overline{B}_d(x,r_m)\subseteq B_d(x^\prime,r_{m^\prime})\right)$.
\item If $(x^\prime,m^\prime)\in X\times\mathbb{N}$ and $\bar{x}\in B_d(x^\prime,r_{m^\prime})$ then $\forall x\in B_d(\bar{x},r_m)\,\big((x,m)<_d(x^\prime,m^\prime)\big)$ for all $m\in\mathbb{N}$ such that $r_m$ is sufficiently close to 0.
\end{enumerate}
\end{lem}

\proof If $(x,m),(x^\prime,m^\prime)\in X\times\mathbb{N}$, $(x,m)<_d(x^\prime,m^\prime)$, $\bar{x}\in X$, $d(x,\bar{x})\!\le\!r_m$ then
$$d(x^\prime,\bar{x})\le d(x^\prime,x)+d(x,\bar{x})<(r_{m^\prime}-r_m)+r_m=r_{m^\prime}.$$
If $(x^\prime,m^\prime)\in X\times\mathbb{N}$, $\bar{x}\in X$, $d(x^\prime,\bar{x})<r_{m^\prime}$, and the natural number $m$ is chosen so that $2r_m<r_{m^\prime}-d(x^\prime,\bar{x})$, then, whenever $x\in X$ and $d(\bar{x},x)< r_m$, we will have
$$d(x^\prime,x)\le d(x^\prime,\bar{x})+d(\bar{x},x)<d(x^\prime,\bar{x})+r_m<r_{m^\prime}-r_m,$$
thus $(x,m)<_d(x^\prime,m^\prime)$.\qed

\begin{thm}\label{T:UVcomp->exretas}
Let $(X,d,\alpha)$ be a semi{\hyp}computable metric space and the set $L$ be recursively enumerable.
If $f$ is $\,({\mathcal U},{\mathcal V})${\hyp}computable then there exists a recursively enumerable topological $(\alpha,\beta)${\hyp}approximation system for $f$. 
\end{thm}

\proof The semi{\hyp}computability of $(X,d,\alpha)$ implies the recursive enumerability of the set
$$\{\,(k,m,k^\prime,m^\prime)\in K\times\mathbb{N}\times K\times\mathbb{N}\ |\ (\alpha(k),m)<_d(\alpha(k^\prime),m^\prime)\,\}.$$
To satisfy the assumptions of Theorem \ref{T:exrealphabetaas<-UVcomp}, we set $H$ to consist of all 
$(k_1,m_1,k_2,m_2,k,m)$ in $K\times\mathbb{N}\times K\times\mathbb{N}\times K\times\mathbb{N}$ such that
$$(\alpha(k),m)<_d(\alpha(k_1),m_1),\ (\alpha(k),m)<_d(\alpha(k_2),m_2).$$
The recursive enumerability of $H$ is clear. Now let $k_1,k_2\in K$ and  $m_1,m_2\in\mathbb{N}$. If $(k_1,m_1,k_2,m_2,k,m)\in H$, then, by the statement (a) of Lemma \ref{L:forminclprop}, the set $U_{\langle k,m \rangle}$ is a subset of each of the sets $U_{\langle k_1,m_1 \rangle}$ and $U_{\langle k_2,m_2 \rangle}$, hence it is a subset of $U_{\langle k_1,m_1\rangle}\cap U_{\langle k_2,m_2\rangle}$. Therefore
$$\bigcup\left\{U_{\langle k,m\rangle}\,\left|\,(k_1,m_1,k_2,m_2,k,m)\in H\right.\right\}\subseteq U_{\langle k_1,m_1\rangle}\cap U_{\langle k_2,m_2\rangle}.$$
For the proof of the converse inclusion, suppose $\bar{x}\in U_{\langle k_1,m_1\rangle}\cap U_{\langle k_2,m_2\rangle}$. Then, by the statement (b) of Lemma~\ref{L:forminclprop}, there exists some $m\in\mathbb{N}$ such that
$$\forall x\in B_d(\bar{x},r_m)\big((x,m)<_d(\alpha(k_1),m_1)\ \&\ (x,m)<_d(\alpha(k_2),m_2))\big).$$
If we take a number $k\in K$ so that $\alpha(k)\in B_d(\bar{x},r_m)$, then we will have $\bar{x}\in U_{\langle k,m\rangle}$ and $(k_1,m_1,k_2,m_2,k,m)\in H$.\qed

\begin{cor}\label{comp<->ex_re_tas}
Under the assumptions of Theorem \ref{T:UVcomp->exretas}, the function $f$ is $\,({\mathcal U},{\mathcal V})${\hyp}computable iff there exists a recursively enumerable topological $(\alpha,\beta)${\hyp}approximation system for~$f$.
\end{cor}

The next example gives a simple characterization of the computability of real functions (similar applications can be done for computability of functions with domains and ranges in other mathematically significant spaces).\footnote{Computability of real functions is usually defined via the Cauchy representation of the real numbers, but this representation is computably equivalent to the topological one (cf. \cite[Theorem 8.1.4]{ca}; see also Subsection \ref{alphabeta-UV} of this paper).}

\begin{example}\label{simplechar}
{\em Let $p,q$ be positive integers, $X=\mathbb{R}^p$, $Y=\mathbb{R}^q$. Let $d,e$ be the Euclidean metrics in $X$ and $Y$, respectively, or the metrics defined by
\begin{align*}
d((x_1,\ldots,x_p),(x_1^\prime,\ldots,x_p^\prime))=&\,\max(|x_1-x_1^\prime|,\ldots,|x_p-x_p^\prime|),\\
e((y_1,\ldots,y_q),(y_1^\prime,\ldots,y_q^\prime))=&\,\max(|y_1-y_1^\prime|,\ldots,|y_q-y_q^\prime|).
\end{align*}
Let $K$ and $L$ be recursively enumerable, $\mathrm{rng}(\alpha)=\mathbb{Q}^p$, $\mathrm{rng}(\beta)=\mathbb{Q}^q$, and $\alpha$, $\beta$ be computable. Then $f$ is $\,({\mathcal U},{\mathcal V})${\hyp}computable iff a recursively enumerable subset $T$ of $\mathbb{Q}^p\times\mathbb{N}\times\mathbb{Q}^q\times\mathbb{N}$ exists such that
\begin{equation}\label{E:simplechar}
\forall x\in E\ \forall b\in\mathbb{Q}^q\,\forall n\in\mathbb{N}\,\big(e(b,f(x))<r_n\Leftrightarrow\exists a,m\,\big((a,m,b,n)\in T\ \&\,d(a,x)<r_m\big)\big).
\end{equation}
This can be proved as follows. By Corollary~\ref{comp<->ex_re_tas}, the function $f$ is $\,({\mathcal U},{\mathcal V})${\hyp}computable iff 
there exists a recursively enumerable topological $(\alpha,\beta)${\hyp}approximation system for $f$, i.e. iff there exists a recursively enumerable subset $S$ of~$K\times\mathbb{N}\times L\times\mathbb{N}$ such that
\begin{equation}\label{E:simplechar'}
\forall x\!\in\!E\ \forall l\!\in\!L\ \forall n\!\in\!\mathbb{N}\big(e(\beta(l),f(x))\!<\!r_n\Leftrightarrow\!\exists k,m\big((k,m,l,n)\!\in\!S\ \&\ d(\alpha(k),x)\!<\!r_m\big)\big).
\end{equation}
It is easy to see that such a set $S$ exists iff a recursively enumerable subset $T$ of $\mathbb{Q}^p\times\mathbb{N}\times\mathbb{Q}^q\times\mathbb{N}$ exists with the property (\ref{E:simplechar}). Indeed, if $S$ is a recursively enumerable subset of~$K\times\mathbb{N}\times L\times\mathbb{N}$ satisfying (\ref{E:simplechar'}), and we set
$$T=\{(\alpha(k),m,\beta(l),n)\,|\,(k,m,l,n)\in S\},$$
then $T$ is a recursively enumerable subset of $\mathbb{Q}^p\times\mathbb{N}\times\mathbb{Q}^q\times\mathbb{N}$ with the property (\ref{E:simplechar}), and, whenever $T$ is a  recursively enumerable subset of $\mathbb{Q}^p\times\mathbb{N}\times\mathbb{Q}^q\times\mathbb{N}$ satisfying (\ref{E:simplechar}), the set
$$S=\{(k,m,l,n)\in K\times\mathbb{N}\times L\times\mathbb{N}\,|\,(\alpha(k),m,\beta(l),n)\in T\}$$
is recursively enumerable and satisfies (\ref{E:simplechar'}).}
\end{example}

 \subsection{Metric approximation systems}\label{S:mas}
Another kind of approximation systems will be introduced now.
\begin{defi}\label{apprsyst}
A {\em metric $(\alpha,\beta)${\hyp}approximation system for the function $f$} is any subset~$S$ of $K\times\mathbb{N}\times L\times\mathbb{N}$ satisfying the following two conditions for any $x\in E$:
\begin{enumerate}[label=\({\alph*}]
\item $\forall(k,m,l,n)\in S\left(d(\alpha(k),x)<r_m\ \Rightarrow\ e(\beta(l),f(x))<r_n\right)$.
\item $\forall n\in\mathbb{N}\,\exists m\in\mathbb{N}\,\forall k\in K\left(d(\alpha(k),x)<r_m\ \Rightarrow\ \exists l\big((k,m,l,n)\in S\big)\right)$.
\end{enumerate}
\end{defi}

The notion introduced in the above definition can be regarded as a slight generalization of a notion introduced in \cite{ed}. If $K\subseteq X$, $\alpha=\mathrm{id}_K$, $L\subseteq Y$, $\beta=\mathrm{id}_L$ then the metric $(\alpha,\beta)${\hyp}approximation systems for $f$ are exactly the $K,L${\hyp}approximation systems for $f$ in the sense of \cite{ed}. The next example and certain considerations going after it concern the relation of the arbitrary metric $(\alpha,\beta)${\hyp}approximation systems to this particular instance.

\begin{example}\label{Hat}
{\em Let $S_0\subseteq A\times\mathbb{N}\times B\times\mathbb{N}$, where $A=\mathrm{rng}(\alpha)$, $B=\mathrm{rng}(\beta)$, and let
\begin{equation}\label{hat}
S=\{(k,m,l,n)\in K\times\mathbb{N}\times L\times\mathbb{N}\,|\,(\alpha(k),m,\beta(l),n)\in S_0\}.
\end{equation}
Then $S$ is a metric $(\alpha,\beta)${\hyp}approximation system for $f$ iff $S_0$ is an $A,B${\hyp}approximation system for~$f$ in the sense of \cite{ed}.}
\end{example}

\begin{defi}\label{saturated}
A subset $S$ of $K\times\mathbb{N}\times L\times\mathbb{N}$ will be called {\em saturated} if
$$\forall(k,m,l,n)\in S\,\forall \bar{k}\in K\,\forall \bar{l}\in L\,\big(\alpha(\bar{k})=\alpha(k)\ \&\ \beta(\bar{l})=\beta(l)\ \Rightarrow\ (\bar{k},m,\bar{l},n)\in S\big).$$
\end{defi}

\begin{example}\label{clearly}
{\em A set $S$ defined in the way from Example \ref{Hat} is always saturated, and if the mappings $\alpha$ and $\beta$ are injective then all subsets of $K\times\mathbb{N}\times L\times\mathbb{N}$ are saturated.}
\end{example}

\begin{example}\label{converse}
{\em Let $S$ be a saturated subset of $K\times\mathbb{N}\times L\times\mathbb{N}$, and let
\begin{equation}\label{-hat}
S_0=\{\alpha(k),m,\beta(l),n)\,|\,(k,m,l,n)\in S\}.
\end{equation}
The assumption that $S$ is saturated and the equality (\ref{-hat}) imply the equality~(\ref{hat}). Therefore $S$ is a metric $(\alpha,\beta)${\hyp}approximation system for $f$ iff $S_0$ is a $A,B${\hyp}approximation system for~$f$ in the sense of \cite{ed} with $A=\mathrm{rng}(\alpha)$, $B=\mathrm{rng}(\beta)$.}
\end{example}

\begin{remark}\label{sat-needed}
{\em The assumption that $S$ is saturated cannot be omitted in Example~\ref{converse}. For instance, let both $\mathbf{X}$ and $\mathbf{Y}$ be the set of the non{\hyp}negative real numbers with the usual metric, $K=L=\mathbb{Q}$, $\alpha=\beta$, $\alpha(k)=|k|$ for any $k\in K$, and $f(x)=x$ for all $x\in X$. Let $S=\{(k,m,l,n)\in K\times\mathbb{N}\times L\times\mathbb{N}\,|\,k=l\ge 0, m=n\}$, and $S_0$ be defined by means of~(\ref{-hat}). Then $S_0=S$, and $S_0$ is a $\mathrm{rng}(\alpha),\mathrm{rng}(\beta)${\hyp}approximation system for $f$ in the sense of~\cite{ed}. However, $S$ is not a metric $(\alpha,\beta)${\hyp}approximation system for $f$, since condition (b) of Definition \ref{apprsyst} is violated for~$S$.} 
\end{remark}

A metric $(\alpha,\beta)${\hyp}approximation system for $f$ could happen to be not saturated. Here is an example.

\begin{example}\label{non-sat}
{\em Let again both $\mathbf{X}$ and $\mathbf{Y}$ be the set of the non{\hyp}negative real numbers with the usual metric, but now both $K$ and $L$ be the set $\mathbb{N}^2$, and let $\alpha(p,q)=\beta(p,q)=\frac{p}{q+1}$ for any $(p,q)\in\mathbb{N}^2$. Let $E$ be the set of the positive irrational numbers, and let $f(x)=x/2$ for any $x\in E$. Then the set $S=\{((p,q),m,(\lfloor p/2\rfloor,q),n)\,|\,(p,q)\in\mathbb{N}^2,\,q\ge n,\,m\ge n\}$ is a metric $(\alpha,\beta)${\hyp}approximation system for $f$. This set is not saturated. For instance, $((1,0),0,(0,0),0)\in S$, but $((2,1),0,(0,0),0)\not\in S$, although $\alpha(1,0)=\alpha(2,1)$.}
\end{example}

\begin{remark}
{\em If $S$ is a metric $(\alpha,\beta)${.\hyp}approximation system for $f$ then
$$\left\{(k,m,l,n)\,\left|\,k\in K\ \&\ l\in L\ \&\ \exists k^\prime,l^\prime\big((k^\prime,m,l^\prime,n)\in S\ \&\ \alpha(k)=\alpha(k^\prime)\ \&\ \beta(l)=\beta(l^\prime)\big)\right.\right\}$$
is a saturated $(\alpha,\beta)${\hyp}approximation system for $f$.}
\end{remark}

\begin{prop}\label{continuous}
A metric $(\alpha,\beta)${\hyp}approximation system for $f$ exists iff $f$ is continuous. If $f$ is continuous then the set of all $(k,m,l,n)$ in $K\times\mathbb{N}\times K\times\mathbb{N}$ satisfying the condition \ref{max} is a saturated metric $(\alpha,\beta)${\hyp}approximation system for $f$.
\end{prop}

\proof To prove that the existence of a metric $(\alpha,\beta)${\hyp}approximation system for the function~$f$ implies its continuity, suppose $S$ is a metric $(\alpha,\beta)${\hyp}approximation system for~$f$. Let $x\in E$, and $\varepsilon$ be a positive number. We choose a natural number $n$ satisfying the inequality $2r_n<\varepsilon$. By condition~(b) of Definition~\ref{apprsyst}, a natural number $m$ exists such that, whenever $k\in K$ and $d(\alpha(k),x)<r_m$, the quadruple $(k,m,l,n)$ belongs to $S$ for some $l$. Let $m$ be such a natural number, and let $x^\prime$ be any element of $E$ satisfying the inequality $d(x^\prime,x)<r_m/2$. After choosing an element $k$ of $K$, satisfying the inequality $d(\alpha(k),x^\prime)<r_m/2$, we will have both inequalities $d(\alpha(k),x^\prime)<r_m$ and $d(\alpha(k),x)<r_m$. By the second of them, $(k,m,l,n)\in S$ for some $l$. The two inequalities and condition (a) of Definition~\ref{apprsyst} imply the inequalities $e(\beta(l),f(x^\prime))<r_n$ and $e(\beta(l),f(x))<r_n$, hence $e(f(x^\prime),f(x))<2r_n<\varepsilon$. Thus the continuity of $f$ is established. For proving the rest of the statement of the theorem, suppose now $f$ is continuous. Let $S$ be the set of all \mbox{$(k,m,l,n)\in K\times\mathbb{N}\times L\times\mathbb{N}$} which satisfy the condition~(\ref{max}). This set is obviously saturated. We will prove that $S$ is a metric $(\alpha,\beta)${\hyp}approximation system for~$f$. Let $x$ be an arbitrary element of~$E$. Condition~(a) of Definition~\ref{apprsyst} follows immediately from the definition of~$S$. To check condition~(b), suppose $n$ is a natural number. By the continuity of $f$, a positive number $\delta$ exists such that $e(f(x^\prime),f(x))<r_n/2$ for all $x^\prime$ in~$E$ satisfying the inequality $d(x^\prime,x)<\delta$. We choose a natural number $m$ with $2r_m<\delta$ and an element $l$ of $L$ such that $e(\beta(l),f(x))<r_n/2$. Consider now any $k$ in~$K$ satisfying the inequality $d(\alpha(k),x)<r_m$. We will show that $(k,m,l,n)\in S$. To do this, suppose $x^\prime$ is any element of $E$ satisfying the inequality $d(\alpha(k),x^\prime)<r_m$. Then 
$$d(x^\prime,x)\le d(x^\prime,\alpha(k))+d(\alpha(k),x)<2r_m<\delta$$
and consequently $e(f(x^\prime),f(x))<r_n/2$. Since 
$$e(\beta(l),f(x^\prime))\le e(\beta(l),f(x))+e(f(x),f(x^\prime)),$$
we see that $e(\beta(l),f(x^\prime))<r_n$.\qed

\begin{remark}\label{maximal}
{\em If $f$ is continuous then the set of all $(k,m,l,n)\in K\times\mathbb{N}\times L\times\mathbb{N}$ satisfying the condition (\ref{max}) is obviously the maximal metric $(\alpha,\beta)${\hyp}approximation system for $f$. Let us note that, as seen from the above proof, condition~(b) of Definition \ref{apprsyst} is satisfied for this set in a stronger form, namely $l$ does not depend on the choice of $k$. By Remark \ref{maxtop}, the set in question is also the maximal topological $(\alpha,\beta)${\hyp}approximation system for $f$.}
\end{remark}

 \subsection{Metric approximation systems and TTE computability}\label{comp}
In this subsection, the sets $K$ and $L$ will be supposed to be subsets of~$\mathbb{N}$ (hence the metric $(\alpha,\beta)${\hyp}approximation systems for $f$ will be subsets of $\mathbb{N}^4$). Any element $x$ of $X$ will be named (in the sense of TTE) by the functions $u:\mathbb{N}\to K$ such that \mbox{$\forall t\in\mathbb{N}\big(d(\alpha(u(t)),x)<r_t\big)$,} and, similarly, any element $y$ of~$Y$ will be named by the functions $v:\mathbb{N}\to L$ such that $\forall t\in\mathbb{N}\big(e(\beta(v(t)),y)<r_t\big)$. The functions $u$ and $v$ in question will be called {\em$\alpha${\hyp}names} of $x$ and {\em$\beta${\hyp}names} of $y$, respectively. The function $f$ will be called {\em $(\alpha,\beta)${\hyp}computable} if a recursive operator $F$ exists such that $F(u)$ is a $\beta$-name of~$f(x)$, whenever $x\in E$ and $u$ is an $\alpha$-name of $x$.

\begin{remark}
{\em As it is easy to see, the $(\alpha,\beta)${\hyp}computability of $f$ does not depend on the choice of the sequence $r_0,r_1,r_2,\ldots$ Indeed, suppose $r_0^\prime,r_1^\prime,r_2^\prime,\ldots$ is another computable sequence of positive rational numbers, and 0 is an accumulation point also of this sequence. Then there exists a recursive operator transforming all names of the elements of $X$ based on $r_0,r_1,r_2,\ldots$ into their names based on $r_0^\prime,r_1^\prime,r_2^\prime,\ldots$, as well as a recursive operator acting in the opposite direction, and similarly for the names of the elements of $Y$. For instance, the first of these operators will transform each function $u:\mathbb{N}\to K$ into the function $\lambda t.u(\mu s[r_s\le r_t^\prime])$.}
\end{remark}

\begin{remark}
{\em In the case of an often adopted version of the definition for computability in metric spaces, we have $r_t=2^{-t}$ for any $t\in\mathbb{N}$, any element $x$ of $X$ is named by the functions $u:\mathbb{N}\to K$ such that $\forall t,h\in\mathbb{N}\big(d(\alpha(u(t)),\alpha(u(t+h)))<r_t\big)$ and $\lim_{t\to\infty}u(t)=x$, similarly for the elements of $Y$. The TTE computability of $f$ corresponding to this sort of naming of the elements of $X$ and $Y$ is equivalent to $(\alpha,\beta)${\hyp}computability of $f$. The same holds if $r_0,r_1,r_2,\ldots$ is any other computable monotonically decreasing sequence of rational numbers converging to 0.}
\end{remark}

\begin{thm}\label{ex_remas->comp}
If there exists a recursively enumerable metric $(\alpha,\beta)${\hyp}approximation system for the function $f$ then $f$ is $(\alpha,\beta)${\hyp}computable.
\end{thm}

\proof Let $S$ be a recursively enumerable metric $(\alpha,\beta)${\hyp}approximation system for $f$. By the recursive enumerability of $S$, a 5{\hyp}argument primitive recursive function~$\chi$ can be found such that, for all $k,m,l,n\in\mathbb{N}$, the equivalence $$(k,m,l,n)\in S\Leftrightarrow\exists s\in\mathbb{N}\big(\chi(k,m,l,n,s)=0\big)$$ holds. Let $\pi_1,\pi_2,\pi_3$ be unary primitive recursive functions such that
$$\{(\pi_1(t),\pi_2(t),\pi_3(t))\,|\,t\in\mathbb{N}\}=\mathbb{N}^3.$$ Let us define recursive operators $T$ and $F$ as follows:
\begin{equation*}
T(u)(n)=\mu t[\,\chi(u(\pi_1(t)),\pi_1(t),\pi_2(t),n,\pi_3(t))=0\,],\ \ 
F(u)(n)=\pi_2(T(u)(n)).
\end{equation*}
We will prove the $(\alpha,\beta)${\hyp}computability of $f$ by showing that, whenever \mbox{$x\in E$} and $u$ is an $\alpha$-name of $x$, the function $F(u)$ is a $\beta$-name of $f(x)$. Let $u$ be an $\alpha$-name of an element $x$ of~$E$, and let $n\in\mathbb{N}$. Making use of condition (b) of Definition~\ref{apprsyst}, we choose a natural number~$m$ such that, whenever $k\in K$ and $d(\alpha(k),x)<r_m$, the quadruple $(k,m,l,n)$ belongs to~$S$ for some $l\in L$. Since $u(m)\in K$ and \mbox{$d(\alpha(u(m)),x)<r_m$,} there exists \mbox{$l\in L$} such that $(u(m),m,l,n)\in S$, hence \mbox{$\chi(u(m),m,l,n,s)=0$} for some $s\in\mathbb{N}$. It follows from here that $n$ belongs to $\mathrm{dom}(T(u))$, and consequently $n$ belongs also to $\mathrm{dom}(F(u))$. After setting $T(u)(n)=i_0$, $\pi_1(i_0)=t_0$, we will have the equality
$$\chi(u(t_0),t_0,F(u)(n),n,\pi_3(i_0))=0,$$
and it shows that $(u(t_0),t_0,F(u)(n),n)\in S$. From here, making use of the inequality $d(\alpha(u(t_0)),x)<r_{t_0}$ and condition (a) of Definition~\ref{apprsyst}, we conclude that
$$e(\beta(F(u)(n)),f(x))<r_n.$$
Since this reasoning was done for an arbitrary natural number $n$, we thus proved that $F(u)$ is really a $\beta$-name of~$f(x)$.\qed

\begin{remark}
{\em Making use of Examples \ref{Hat}, \ref{clearly} and \ref{converse}, we see that the theorem in \cite{ed} corresponding to Theorem \ref{ex_remas->comp} is actually its particular instance, when 
$r_t=\frac{1}{t+1}$ for all $t\in\mathbb{N}$ and there  exists a saturated recursively enumerable metric $(\alpha,\beta)${\hyp}approximation system for~$f$.}
\end{remark}

\begin{example}\label{no_satur}
{\em It is possible that a recursively enumerable metric $(\alpha,\beta)${\hyp}approximation system for~$f$ exists, but there exists no saturated such one. Let $$X=Y=\{0,1\},\ d(0,1)=e(0,1)=1,\ K=L=\mathbb{N},$$ let $\alpha(0)=0$, $\alpha(k)=1$ for any $k\in\mathbb{N}\setminus\{0\}$, $\beta^{-1}(0)$ be not recursively enumerable, $f(0)=0$, $f(1)=1$, and let $r_t\le 1$ for any $t\in\mathbb{N}$. If the natural numbers $l_0$ and $l_1$ are such that $\beta(l_0)=0$ and $\beta(l_1)=1$ then the union of the sets $\{0\}\!\times\!\mathbb{N}\!\times\!\{l_0\}\!\times\!\mathbb{N}$ and \mbox{$(\mathbb{N}\!\setminus\!\{0\})\!\times\!\mathbb{N}\!\times\!\{l_1\}\!\times\!\mathbb{N}$} is a recursively enumerable metric $(\alpha,\beta)${\hyp}approximation system for $f$. However, no saturated recursively enumerable metric $(\alpha,\beta)${\hyp}approximation system for $f$ exists. Indeed, let $S$ be a saturated metric $(\alpha,\beta)${\hyp}approximation system for $f$. Then it is easy to prove that \mbox{$\beta^{-1}(0)=\{l\,|\,(0,0,l,0)\in S\}$,} and therefore $S$ cannot be recursively enumerable.} 
\end{example}

Under an additional assumption, a converse of Theorem \ref{ex_remas->comp} also holds.

\begin{thm}\label{comp->ex_remas}
Let $(X,d,\alpha)$ and $(Y,e,\beta)$ be semi{\hyp}computable metric spaces. If the function $f$ is $(\alpha,\beta)${\hyp}computable then there exists a saturated recursively enumerable metric $(\alpha,\beta)${\hyp}approximation system for $f$.
\end{thm}

\proof The semi{\hyp}computability of $(X,d,\alpha)$ and $(Y,e,\beta)$ implies the recursive enumerability of the sets $K$, $L$ and of the sets
\begin{gather}
\left\{(k,k^\prime,t)\in K^2\times\mathbb{N}\,\left|\ d(\alpha(k),\alpha(k^\prime))<r_t/2\right.\right\},\label{a}\\
\left\{(l,l^\prime,t)\in L^2\times\mathbb{N}\,\left|\ e(\beta(l),\beta(l^\prime))<r_t/2\right.\right\}\ \label{b}
\end{gather}
Let $F$ be a recursive operator such that $F$ transforms unary partial functions in $\mathbb{N}$ into unary partial functions in $\mathbb{N}$ and, whenever $u$ is an $\alpha$-name of an element $x$ of~$E$, the function $F(u)$ is a $\beta$-name of $f(x)$. Let $S$ be the set of all quadruples \mbox{$(k,m,l,n)\in K\times\mathbb{N}\times L\times\mathbb{N}$} such that, for some $s,p\in\mathbb{N}$ satisfying the inequalities $\min\{r_0,r_1,\ldots r_s\}\ge 2r_m$, $r_p\le r_n/2$ and some function $u^\circ:\{0,1,\ldots,s\}\to K$ the following holds:
\begin{gather}
d(\alpha(u^\circ(t)),\alpha(k))<r_t/2,\ \ t=0,1,\ldots,s,\label{d}\\
p\in\mathrm{dom}(F(u^\circ)),\label{in}\\
F(u^\circ)(p)\in L,\ \ e(\beta(F(u^\circ)(p)),\beta(l))<r_n/2\label{me}.
\end{gather}
The set $S$ is obviously saturated. It is recursively enumerable due to the recursiveness of the operator $F$ and the recursive enumerability of the sets $K$, $L$ and the sets (\ref{a}) and~(\ref{b}). We will show that $S$ is a metric $(\alpha,\beta)${\hyp}approximation system for~$f$. 
Let $x\in E$. To verify condition (a) of Definition \ref{apprsyst}, suppose that $(k,m,l,n)\in S$ and $d(\alpha(k),x)<r_m$. Let $s,p$ and $u^\circ$ be two natural numbers and a function with the properties from the above definition of the set $S$. For any $t\in\{0,1,\ldots,s\}$, we have
\begin{equation*}
d(\alpha(u^\circ(t)),x)\le d(\alpha(u^\circ(t)),\alpha(k))+d(\alpha(k),x)<r_t/2+r_m\le r_t.
\end{equation*}
It follows from here that $u^\circ$ can be extended to an $\alpha$-name~$u$ of $x$, and then $F(u)$ will be a $\beta$-name of $f(x)$. By condition (\ref{in}) and the continuity of the operator~$F$, the equality
\begin{equation}\label{eq-restr}
F(u)(p)=F(u^\circ)(p)
\end{equation}
holds. Therefore, making use also of condition (\ref{me}), we have
$$e(\beta(l),f(x))\le e(\beta(l),\beta(F(u^\circ)(p)))+e(\beta(F(u)(p)),f(x))<r_n/2+r_p\le r_n.$$
To verify condition (b), suppose a natural number~$n$ is given. Let \mbox{$u:\mathbb{N}\to K$} be such that $d(\alpha(u(t),x)<r_t/4$ for any $t\in\mathbb{N}$. Clearly $u$ is an $\alpha$-name of~$x$, hence the function $F(u)$ is a $\beta$-name of $f(x)$, and therefore $F(u)$ is total and all its values belong to $L$. Let $l=F(u)(p)$, where $p$ is some natural number satisfying the inequality $r_p\le r_n/2$. By the continuity of the operator~$F$, a natural number $s$ exists such that $u^\circ=u\upharpoonright\{0,1,\ldots,s\}$ satisfies condition~(\ref{in}) and the equality (\ref{eq-restr}), hence \mbox{$l=F(u^\circ)(p)$.} Let $m$ be a natural number such that $r_m\le r_t/4$ for $t=0,1,\ldots,s$, and let $k$ be an arbitrary element of~$K$ satisfying the inequality $d(\alpha(k),x)<r_m$. Then  the inequalities~(\ref{d}) hold, because
$$d(\alpha(u^\circ(t)),\alpha(k))\le d(\alpha(u^\circ(t)),x)+d(x,\alpha(k))<r_t/4+r_m\le r_t/2$$
for any $t\in\{0,1,\ldots,s\}$. Since condition (\ref{me}) is trivially satisfied, we see that $(k,m,l,n)$ belongs to $S$.\qed

\begin{cor}\label{comp<->ex_re_as}
Under the assumptions of Theorem \ref{comp->ex_remas}, the function $f$ is $(\alpha,\beta)${\hyp}computable iff there exists a recursively enumerable metric $(\alpha,\beta)${\hyp}approximation system for $f$.
\end{cor}

\begin{remark}
{\em In effect, the proof of Theorem \ref{comp->ex_remas} shows the truth of the following statement: if the sets \eqref{a} and \eqref{b} are recursively enumerable, and the function $f$ is $(\alpha,\beta)${\hyp}computable, then there exists a saturated recursively enumerable metric $(\alpha,\beta)${\hyp}approximation system for $f$  (note that the recursive enumerability of the sets \eqref{a} and \eqref{b} implies the one of $K$ and $L$). The theorem in \cite{ed} corresponding to Theorem \ref{comp->ex_remas} is actually the instance of this statement with $r_t=\frac{1}{t+1}$ for all $t\in\mathbb{N}$.}
\end{remark}

\begin{remark}
{\em The set $S$ constructed in the proof of Theorem \ref{comp->ex_remas} satisfies condition (b) of Definition~\ref{apprsyst} in the stronger form mentioned in Remark \ref{maximal} (the number $l$ does not depend on the choice of $k$). However, this set is not necessarily the maximal metric $(\alpha,\beta)${\hyp}approximation system for $f$. Moreover, as seen from the next example, the assumptions of Theorem \ref{comp->ex_remas} do not imply that the maximal metric $(\alpha,\beta)${\hyp}approximation system for $f$ is necessarily recursively enumerable.} 
\end{remark}

\begin{example}{\em
Let $X=K=Y=L=\mathbb{N}$, the metrics $d$ and $e$ coincide with the usual metric in $\mathbb{N}$ (i.e. the absolute value of the difference), and $\alpha=\beta=\mathrm{id}_\mathbb{N}$. Let $E$ be such that $\mathbb{N}\setminus E$ is not recursively enumerable, $f$ be the restriction of the constant 0 to $E$, and $S$ be the maximal metric $(\alpha,\beta)${\hyp}approximation system for $f$. Let $m$ and $n$ be natural numbers such that $r_m\le 1$ and $r_n\le 1$. Then, for any $k,l\in\mathbb{N}$, the condition (\ref{max}) is equivalent to the implication $k\in E\Rightarrow l=0$. Therefore $(k,m,1,n)\in S\Leftrightarrow k\not\in E$ for any $k\in\mathbb{N}$, hence $S$ is not recursively enumerable.
}\end{example}

 \subsection{The interrelation between \texorpdfstring{$(\alpha,\beta)$}{(alpha,beta)}- and \texorpdfstring{$({\mathcal U},{\mathcal V})$}{(U,V)}{\hyp}computability in metric spaces}\label{alphabeta-UV}
In this subsection, the situation from Subsection \ref{S:iUVcntas} will be supposed to be present again, thus the question about $({\mathcal U},{\mathcal V})${\hyp}computability of $f$ can be considered as well. In general, the $(\alpha,\beta)${\hyp}computability of the function $f$ and its $({\mathcal U},{\mathcal V})${\hyp}computability can be non{\hyp}equivalent. Here are two examples.

\begin{example}\label{ab-UV}
{\em Let $f$ be the function from Example \ref{no_satur}. Although this function is $(\alpha,\beta)${\hyp}computable, it is not $({\mathcal U},{\mathcal V})${\hyp}computable. Indeed, we have the equalities $$[0]_{\mathcal U}=\{\langle k,m\rangle\,|\,k,m\in\mathbb{N}\ \&\ \alpha(k)<r_m\},\ [f(0)]_{\mathcal V}=\{\langle l,n\rangle\,|\,l,n\in\mathbb{N}\ \&\ \beta(l)<r_n\}.$$ By the first of them and the recursiveness of $\alpha$, the set $[0]_{\mathcal U}$ is recursively enumerable. On the other hand, the set $[f(0)]_{\mathcal V}$ is not recursively enumerable, since if $c$ is a natural number such that $r_c\le 1$ then $\{l\in\mathbb{N}\,|\,\langle l,c\rangle\in[f(0)]_{\mathcal V}\}=\beta^{-1}(0)$. Hence there is no enumeration operator transforming $[0]_{\mathcal U}$ into $[f(0)]_{\mathcal V}$.}
\end{example}

\begin{example}\label{UV-ab}
{\em In the situation from Example \ref{no_eff}, the function $\mathrm{id}_\mathbb{N}$ is not $(\alpha,\beta)${\hyp}computable, although it is $({\mathcal U},{\mathcal V})${\hyp}computable. Indeed, suppose $\mathrm{id}_\mathbb{N}$ is $(\alpha,\beta)${\hyp}computable in this situation. Then there exists a recursive operator $F$ such that, for any $x$ in $\mathbb{N}$, the operator~$F$ transforms each $\alpha$-name of $x$ into some $\beta$-name of $x$. For any natural number~$k$, let $\hat{k}$ be the total constant function with value $k$ in $\mathbb{N}$. Clearly this function is an $\alpha$-name of $\alpha(k)$, therefore $F(\hat{k})$ is a $\beta$-name of $\alpha(k)$. Then $\alpha(k)=F(\hat{k})(0)$ for all $k$ in $\mathbb{N}$, hence $\alpha$ is a recursive function. Since $\mu k[\alpha(k)=x]\in P_x$ for any $x$ in $\mathbb{N}$, we get a recursive function whose value belongs to $P_x$ for any $x$ in $\mathbb{N}$, and this is a contradiction.} 
\end{example}

However, it is a known fact that the topological computability in metric spaces, which is considered in Subsection \ref{S:tasms}, and the metric one considered in Subsection \ref{comp} are equivalent under certain not very restrictive assumptions (cf., for instance, \cite[Theorem 8.1.4]{ca}). For the sake of completeness, we will briefly consider this question.\footnote{The proof of the above-mentioned theorem in the book \cite{ca} is left there as an exercise to the reader. In the paper \cite{ceets}, a theorem of a similar type (namely Theorem 3.7 of that paper) is accompanied by a proof, but, in our opinion, it needs adding some explanation. In \cite[Statement (2) of Theorem~6.2]{crts}, essentially the same result is given as in Lemma \ref{tpl<->mtr} below (with the superfluous requirement about recursiveness of the domains of the enumerations). Unfortunately, it was already late to make substantial changes in the present paper when the paper \cite{crts} appeared.} Let us call {\em$\mathcal U$-names} of an element $x$ of $X$ the total enumerations of the set $[x]_{\mathcal U}$, and let the {\em$\mathcal V$-names} of an element of $Y$ be defined similarly. The $({\mathcal U},{\mathcal V})${\hyp}computability of $f$ is equivalent to the existence of a recursive operator which transforms any $\mathcal U$-name of any element of $E$ into some $\mathcal V$-name of the corresponding value of $f$. For the equivalence of the $(\alpha,\beta)${\hyp}computability of $f$ and its $({\mathcal U},{\mathcal V})${\hyp}computability, it is thus sufficient that recursive operators $\Gamma_{{\mathcal U},\alpha}$ and $\Gamma_{\alpha,{\mathcal U}}$ exist which transform, respectively, the $\mathcal U$-names of the elements of $X$ into their $\alpha$-names and the $\alpha$-names of the elements of $X$ into their $\mathcal U$-names, and similar recursive operators $\Gamma_{{\mathcal V},\beta}$ and $\Gamma_{\beta,{\mathcal V}}$ exist for the $\mathcal V$-names of the elements of $Y$ and their $\beta$-names.

Recursive operators $\Gamma_{{\mathcal U},\alpha}$ and $\Gamma_{{\mathcal V},\beta}$ with the above-mentioned properties always exist. For instance, we may define the operator $\Gamma_{{\mathcal U},\alpha}$ by means the equality
$$\Gamma_{{\mathcal U},\alpha}(u)(m)=\pi_1(\mu i[\pi_2(u(i))=m]),$$
where $\pi_1(\langle s,t\rangle)=s$ and $\pi_2(\langle s,t\rangle)=t$ for all $s,t\in\mathbb{N}$. A sufficient condition for the existence of operators $\Gamma_{\alpha,{\mathcal U}}$ and $\Gamma_{\beta,{\mathcal V}}$ is given by the next lemma, where the relation $<_e$ on $(Y\times\mathbb{N})^2$ is the $(Y,e)${\hyp}analog of the partial ordering $<_d$ on $(X\times\mathbb{N})^2$ introduced in Subsection \ref{S:iUVcntas}, namely $(y,n)<_e(y^\prime,n^\prime)$ iff $e(y^\prime,y)<r_{n^\prime}-r_n$.

\begin{lem}\label{tpl<->mtr}
Let $(X,d,\alpha)$ and $(Y,e,\beta)$ be semi{\hyp}computable metric spaces.
Then there exist recursive operators $\Gamma_{\alpha,{\mathcal U}}$ and $\Gamma_{\beta,{\mathcal V}}$ which transform, respectively, the $\alpha$-names of the elements of $X$ into their $\mathcal U$-names and the $\beta$-names of the elements of $Y$ into their $\mathcal V$-names.
\end{lem}

\proof We will describe the construction of the operator $\Gamma_{\beta,{\mathcal V}}$ (the construction of $\Gamma_{\alpha,{\mathcal U}}$ is similar). By the recursive enumerability of the set
$$\{\,(l,n,l^\prime,n^\prime)\in L\times\mathbb{N}\times L\times\mathbb{N}\ |\ (\beta(l),n)<_e(\beta(l^\prime),n^\prime)\,\},$$
a 5{\hyp}argument primitive recursive function $\chi$ exists such that a quadruple $(l,n,l^\prime,n^\prime)$ of natural numbers belongs to this set iff $\chi(l,n,l^\prime,n^\prime,s)=0$ for some $s$ in $\mathbb{N}$. Let $\pi_1,\pi_2,\pi_3,\pi_4$ be unary primitive recursive functions such that
$\{(\pi_1(t),\pi_2(t),\pi_3(t),\pi_4(t))\,|\,t\in\mathbb{N}\}=\mathbb{N}^4$. We set
\begin{gather*}
\Lambda(v)(t)=\chi(v(\pi_1(t)),\pi_1(t),\pi_2(t),\pi_3(t),\pi_4(t)),\ \ \Delta(v)(p)=\mu q\left[\sum_{t=0}^q\overline{\mathrm{sg}}\,\Lambda(v)(t)>p\right],\\
\Gamma_{\beta,{\mathcal V}}(v)(p)=\langle\pi_2(\Delta(v)(p)),\pi_3(\Delta(v)(p))\rangle.
\end{gather*}
The recursiveness of $\Gamma_{\beta,{\mathcal V}}$ is clear. Suppose now $v$ is a $\beta$-name of an element $y$ of $Y$, and $v^\prime=\Gamma_{\beta,{\mathcal V}}(v)$. We will show that $v^\prime$ is a $\mathcal V$-name of $y$, i.e. $\mathrm{dom}(v^\prime)=\mathbb{N}$ and $\mathrm{rng}(v^\prime)=[y]_{\mathcal V}$. To do this, we note that, evidently, for any $p$ and $q$ in $\mathbb{N}$, the equality $\Delta(v)(p)=q$ holds iff $\Lambda(v)(q)=0$ and there are exactly $p$ natural numbers $t$ less than $q$ such that $\Lambda(v)(t)=0$. We note also that
\begin{equation}\label{y_V}
[y]_{\mathcal V}=\{\langle\pi_2(t),\pi_3(t)\rangle\,|\,t\in\mathbb{N},\,\Lambda(v)(t)=0\}.
\end{equation}
Indeed, let $j\in[y]_{\mathcal V}$. Then $j\in J$ and $y\in V_j$. We have $j=\langle l^\prime,n^\prime)$ for some $l^\prime\in L$, $n^\prime\in\mathbb{N}$ such that $V_j=B_e(\beta(l^\prime),r_{n^\prime})$. By the $(Y,e)${\hyp}analog of statement (b) in Lemma \ref{L:forminclprop}, we may choose a natural number $n$ such that $(\tilde{y},n)<_e(\beta(l^\prime),n^\prime)$ for any $\tilde{y}\in B_e(y,r_n)$. Since $v(n)\in L$ and $e(\beta(v(n)),y)<r_n$, we get that $(\beta(v(n)),n)<_e(\beta(l^\prime),n^\prime)$, hence
$\chi(v(n),n,l^\prime,n^\prime,s)=0$ for some $s$ in $\mathbb{N}$. If $t$ is a natural number such that $(\pi_1(t),\pi_2(t),\pi_3(t),\pi_4(t))=(n,l^\prime,n^\prime,s)$ then $\Lambda(v)(t)=0$ and $j=\langle\pi_2(t),\pi_3(t)\rangle$. Conversely, suppose $t$ is a natural number such that $\Lambda(v)(t)=0$. Let us set $n=\pi_1(t)$, $l^\prime=\pi_2(t)$, $n^\prime=\pi_3(t)$. Then $\chi(v(n),n,l^\prime,n^\prime,\pi_4(t))=0$, hence $v(n),l^\prime\in L$ and $(\beta(v(n)),n)<_e(\beta(l^\prime),n^\prime)$. Since $y\in B_e(\beta(v(n)),r_n)$, the $(Y,e)${\hyp}analog of statement (a) in Lemma \ref{L:forminclprop} yields that $y\in B_e(\beta(l^\prime),r_{n^\prime})=U_{\langle l^\prime,n^\prime\rangle}$, and therefore $\langle\pi_2(t),\pi_3(t)\rangle=\langle l^\prime,n^\prime\rangle\in[y]_{\mathcal V}$. Thus the equality (\ref{y_V}) is established. Obviously $\mathrm{dom}(\Delta(v))$ is an initial segment of $\mathbb{N}$, and $\mathrm{rng}(\Delta(v))=\{t\in\mathbb{N}\,|\,\Lambda(v)(t)=0\}$. Due to the fact that $[y]_{\mathcal V}$ is an infinite set and to the equality (\ref{y_V}), there exist infinitely many $t$ with $\Lambda(v)(t)=0$. Hence $\mathrm{dom}(\Delta(v))=\mathbb{N}$, and therefore $\mathrm{dom}(v^\prime)=\mathbb{N}$ too. Making use of the equality (\ref{y_V}) again, we see that $\mathrm{rng}(v^\prime)=[y]_{\mathcal V}$.\qed

\begin{cor}\label{C:alphabeta<->UV}
If $(X,d,\alpha)$ and $(Y,e,\beta)$ are semi{\hyp}computable metric spaces then the $(\alpha,\beta)${\hyp}computability of the function $f$ and its $({\mathcal U},{\mathcal V})${\hyp}computability are equivalent. 
\end{cor}

By applying Corollaries \ref{comp<->ex_re_tas}, \ref{comp<->ex_re_as} and \ref{C:alphabeta<->UV} we get the following result.

\begin{cor}\label{C:equivcond}
If $(X,d,\alpha)$ and $(Y,e,\beta)$ are semi{\hyp}computable metric spaces then the following conditions are equivalent:
\begin{enumerate}[label=\({\alph*}]
\item The function $f$ is ${\mathcal U,\mathcal V}${\hyp}computable.
\item The function $f$ is $(\alpha,\beta)${\hyp}computable.
\item There exists a recursively enumerable topological $(\alpha,\beta)${\hyp}approximation system for $f$.
\item There exists a recursively enumerable metric $(\alpha,\beta)${\hyp}approximation system for $f$.
\end{enumerate}
\end{cor}

 \subsection{Transformation of metric approximation systems into topological ones and vice versa}\label{meto}
Corollary~\ref{tas<->cont} and Proposition \ref{continuous} imply the next statement.

\begin{cor}
A topological $(\alpha,\beta)${\hyp}approximation system for $f$ exists iff there exists a metric $(\alpha,\beta)${\hyp}approximation system for $f$.
\end{cor}

The above corollary suggests the problem about constructions for the direct transformation of metric approximation systems into topological ones and vice versa. Such constructions will be given in the next two theorems.

\begin{thm}\label{mt}
Let $S$ be a metric $(\alpha,\beta)${\hyp}approximation system for $f$, and let
$$S^\prime=\left\{(k,m,l^\prime,n^\prime)\in\mathbb{N}^4\,\left|\,l^\prime\in L\ \&\ \exists l,n\big((k,m,l,n)\in S\ \&\ (\beta(l),n)<_e(\beta(l^\prime),n^\prime)\big)\right.\right\}.$$
Then $S^\prime$ is a topological $(\alpha,\beta)${\hyp}approximation system for $f$.
\end{thm}

\proof We have to prove that
\begin{multline*}
\forall x\in E\ \forall(l^\prime,n^\prime)\in L\times\mathbb{N}\big(e(\beta(l^\prime),f(x))<r_{n^\prime}\Leftrightarrow\\
\exists k,m\big((k,m,l^\prime,n^\prime)\in S^\prime\ \&\ d(\alpha(k),x)<r_m\big)\big).
\end{multline*}
Let $x\in E$, $l^\prime\in L$, $n^\prime\in\mathbb{N}$. Suppose first that
$e(\beta(l^\prime),f(x))<r_{n^\prime}$. Then $f(x)\in B_e(\beta(l^\prime),r_{n^\prime})$. By the $Y,d${\hyp}analog of statement (b) in Lemma \ref{L:forminclprop}, there exists a natural number $n$ such that $(y,n)<_e(\beta(l^\prime),n^\prime)$ for any $y\in B_e(f(x),r_n)$. By consecutively using the properties (b) and (a) from Definition~\ref{apprsyst} (making use also of the fact that $\mathrm{rng}(\alpha)$ is dense in~$X$) we may choose $m\in\mathbb{N}$, $k\in K$, $l\in L$ such that $d(\alpha(k),x)<r_m$, $(k,m,l,n)\in S$ and \mbox{$e(\beta(l),f(x))<r_n$.} Then $\beta(l)\in B_e(f(x),r_n)$ and therefore $(\beta(l),n)<_e(\beta(l^\prime),n^\prime)$. Hence \mbox{$(k,m,l^\prime,n^\prime)\in S^\prime\ \&\ d(\alpha(k),x)<r_m$.} Conversely, suppose this conjunction holds for some $k$ and $m$. Then there exist natural numbers $l$ and $n$ such that $(k,m,l,n)\in S$ and \mbox{$(\beta(l),n)<_e(\beta(l^\prime),n^\prime)$.} By the property (a) from Definition~\ref{apprsyst} and the inequality $d(\alpha(k),x)<r_m$, the inequality $e(\beta(l),f(x))<r_n$ holds, hence $f(x)\in B_e(\beta(l),n)$. Since, by the $(Y,e)${\hyp}analog of statement~(a) in Lemma \ref{L:forminclprop}, $B_e(\beta(l),n)$ is a subset of $B_e(\beta(l^\prime),n^\prime)$, it follows that $f(x)\in B_e(\beta(l^\prime),n^\prime)$ and therefore $e(\beta(l^\prime),f(x))<r_{n^\prime}$.\qed

\begin{thm}\label{tm}
Let $S$ be a topological $(\alpha,\beta)${\hyp}approximation system for $f$, and let
$$S^\prime=\left\{(k^\prime,m^\prime,l,n)\in\mathbb{N}^4\,\left|\,k^\prime\in K\ \&\ \exists k,m\big((k,m,l,n)\in S\ \&\ (\alpha(k^\prime),m^\prime)<_d(\alpha(k),m)\big)\right.\right\}.$$
Then $S^\prime$ is a metric $(\alpha,\beta)${\hyp}approximation system for $f$.
\end{thm}

\proof Let $x\in E$. We have to prove that
\begin{gather}
\forall(k^\prime,m^\prime,l,n)\in S^\prime\left(d(\alpha(k^\prime),x)<r_{m^\prime}\ \Rightarrow\ e(\beta(l),f(x))<r_n\right),\label{1}\\
\forall n\in\mathbb{N}\,\exists m^\prime\in\mathbb{N}\,\forall k^\prime\in K\left(d(\alpha(k^\prime),x)<r_{m^\prime}\ \Rightarrow\ \exists l\big((k^\prime,m^\prime,l,n)\in S^\prime\big)\right).\label{2}
\end{gather}
For proving (\ref{1}), suppose $(k^\prime,m^\prime,l,n)\in S^\prime$ and $d(\alpha(k^\prime),x)<r_{m^\prime}$. Then $x\in B_d(\alpha(k^\prime),r_{m^\prime})$ and there exist $k$ and $m$ such that $(k,m,l,n)\in S$ and $(\alpha(k^\prime),m^\prime)<_d(\alpha(k),m)$. By the statement~(a) in Lemma \ref{L:forminclprop}, $B_d(\alpha(k^\prime),r_{m^\prime})\subseteq B_d(\alpha(k),r_m)$, hence $x\in B_d(\alpha(k),r_m)$. Thus $d(\alpha(k),x)<r_m$, therefore $e(\beta(l),f(x))<r_n$. For the proof of (\ref{2}), let a natural number $n$ be given. Making use of the fact that $\mathrm{rng}(\beta)$ is dense in $Y$, we choose some $l$ in $\mathbb{N}$ satisfying the inequality $e(\beta(l),f(x))<r_n$. Then there exist $k$ and $m$ such that $(k,m,l,n)\in S$ and $d(\alpha(k),x)<r_m$. Since $x\in B_d(\alpha(k),r_m)$, the statement~(b)  in Lemma \ref{L:forminclprop} implies the existence of a natural number $m^\prime$ such that $\forall\bar{x}\in B_d(x,m^\prime)\big((\bar{x},m^\prime)<_d(\alpha(k),m)\big)$. Now let $k^\prime$ be any number from $K$ satisfying the inequality $d(\alpha(k^\prime),x)<r_{m^\prime}$. Then $\alpha(k^\prime)\in B_d(x,m^\prime)$, hence $(\alpha(k^\prime),m^\prime)<_d(\alpha(k),m)$, and therefore $(k^\prime,m^\prime,l,n)\in S^\prime$.\qed

\begin{remark}
{\em As seen from Corollary \ref{C:equivcond}, the existence of a recursively enumerable topological $(\alpha,\beta)${\hyp}approximation system for $f$ is equivalent to the existence of a recursively enumerable metric $(\alpha,\beta)${\hyp}approximation system for $f$ under the assumptions made in the beginning of Subsection~\ref{S:iUVcntas} and the assumption that $(X,d,\alpha)$ and $(Y,e,\beta)$ are semi{\hyp}computable metric spaces. Theorems \ref{mt} and \ref{tm} can be used to give a more direct proof of this equivalence, because then, both in the situation from Theorem \ref{mt} and in the one from Theorem \ref{tm}, the recursive enumerability of $S$ implies the recursive enumerability of~$S^\prime$. This allows the using of Corollary \ref{comp<->ex_re_as} to be avoided in the proof of Corollary \ref{C:equivcond}.}
\end{remark}

\section*{Acknowledgement}
The author would like to express sincere thanks to the referees for numerous very helpful remarks and suggestions.


\begin{thebibliography}{GrWe}

\bibitem[Br]{cts} 
Vasco Brattka.
\newblock Computability over topological structures.
\newblock In: S. Barry Cooper and~Sergey S. Goncharov, editors, Computability and Models, pp.~93--136. Kluwer Academic Publishers, New York, 2003.

\bibitem[GrWe]{cm} 
Tanja Grubba, Klaus Weihrauch.
\newblock On computable metrization.
\newblock Electron. Notes Theor. Comput. Sci. {\bf 167} (2007) 345--364.\\
\url{http://dx.doi.org/10.1016/j.entcs.2006.08.020}

\bibitem[He]{emsrr} 
Armin Hemmerling.
\newblock Effective metric spaces and representations of the reals.
\newblock Theoret. Comput. Sci. {\bf 284} (2002) 347--372\\
\url{http://dx.doi.org/10.1016/S0304-3975(01)00093-7}

\bibitem[KoKu]{ceets} 
Margarita Korovina, Oleg Kudinov.
\newblock Towards computability over effectively enumerable topological spaces.
\newblock Electron. Notes Theor. Comput. Sci. {\bf 221} (2008) 115--125.\\
\url{http://dx.doi.org/10.1016/j.entcs.2008.12.011}

\bibitem[Ro]{trfec} 
Hartley Rogers, Jr.
\newblock Theory of Recursive Functions and Effective Computability.
\newblock McGraw-Hill, 1967.

\bibitem[Sk]{ed} 
Dimiter Skordev.
\newblock An epsilon-delta characterization of a certain TTE computability notion.
\newblock arXiv:1207.7270 [math.LO], 2012.\\
\url{http://arxiv.org/abs/1207.7270}

\bibitem[We93]{cms}
Klaus Weihrauch.
\newblock Computability on computable metric spaces.
\newblock Theoret. Comput. Sci. {\bf 113} (1993) 191--210.\\
\url{http://dx.doi.org/10.1016/0304-3975(93)90001-A}

\bibitem[We0]{ca} 
Klaus Weihrauch.
\newblock Computable Analysis. An Introduction.
\newblock Berlin/Heidelberg, Springer-Verlag, 2000.

\bibitem[We8]{mr}
Klaus Weihrauch.
\newblock The computable multi-functions on multi-represented sets are closed under programming. \newblock J. Univers. Comput. Sci. {\bf 14} (2008) 801–-844.\\
\url{http://dx.doi.org/10.3217/jucs-014-06-0801}

\bibitem[We13]{crts}
Klaus Weihrauch.
\newblock Computably regular topological spaces.
\newblock Log. Meth. Comput. Sci. {\bf 9(3:5)} (2013) 1--24\\
\url{http://www.lmcs-online.org/ojs/viewarticle.php?id=963}

\bibitem[WeGr]{ect}
Klaus Weihrauch, Tanja Grubba.
\newblock Elementary computable topology.
\newblock J. Univers. Comput. Sci. {\bf 15} (2009) 1381--1422.\\
\url{http://dx.doi.org/10.3217/jucs-015-06-1381}
\end{thebibliography}
\end{document}